\documentclass[usenatbib]{mnras}

\usepackage{newtxtext,newtxmath}

\usepackage{graphicx}
\usepackage{amsmath}
\usepackage{amssymb}

\usepackage[T1]{fontenc}
\usepackage[normalem]{ulem}
\usepackage{academicons}
\usepackage{hyperref}
\usepackage{xcolor}
\UseRawInputEncoding

\newcommand{\orcid}[1]{\href{https://orcid.org/#1}{\textcolor[HTML]{A6CE39}{\aiOrcid}}}

\DeclareRobustCommand{\VAN}[3]{#2}

\title[21cm signal from Cosmic Dawn]{The role of Pop III stars and early black holes in the 21cm signal from Cosmic Dawn}

\author[E. M. Ventura et al.]{
Emanuele M. Ventura,$^{1,2}$\thanks{E-mail: eventura@student.unimelb.edu.au}
Alessandro Trinca$^{3,4}$,
Raffaella Schneider$^{3,4,5,6}$, 
Luca Graziani$^{3,5}$ 
\newauthor
Rosa Valiante$^{4,5}$
and J. Stuart B. Wyithe$^{1,2}$
\\
$^{1}$School of Physics, University of Melbourne, Parkville, Victoria, Australia\\
$^{2}$ARC Centre of Excellence for All Sky Astrophysics in 3 Dimensions (ASTRO 3D)\\
$^{3}$Dipartimento di Fisica, Universita' di Roma La Sapienza, Piazzale Aldo Moro 2, 00185 Roma, Italy\\
$^{4}$INFN, Sezione Roma1, Dipartimento di Fisica, Universita' di Roma La Sapienza, Piazzale Aldo Moro 2, 00185, Roma, Italy\\
$^{5}$INAF/Osservatorio Astronomico di Roma, Via di Frascati 33, 00040 Monte Porzio Catone, Italy \\
$^{6}$Sapienza School for Advanced Studies, Viale Regina Elena 291, 00161 Roma, Italy
}

\date{Accepted XXX. Received YYY; in original form ZZZ}

\pubyear{2022}

\begin{document}
\label{firstpage}
\pagerange{\pageref{firstpage}--\pageref{lastpage}}
\maketitle

\begin{abstract} 
Modeling the 21cm global signal from the Cosmic Dawn is challenging due to the many poorly constrained physical processes that come into play. We address this problem using the semi-analytical code "Cosmic Archaeology Tool" (\textsc{cat}). \textsc{cat} follows the evolution of dark matter halos tracking their merger history and provides an ab initio description of their baryonic evolution, starting from the formation of the first (Pop III) stars and black holes (BHs) in mini-halos at z > 20. The model is anchored to observations of galaxies and AGN at z < 6 and predicts a reionization history consistent with constraints. In this work we compute the evolution of the mean global 21cm signal between $4\leq z \leq 40$ based on the rate of formation and emission properties of stars and accreting black holes. We obtain an absorption profile with a maximum depth $\delta {\rm T_b} = -95$ mK at $z \sim 26.5$ (54 MHz). This feature is quickly suppressed turning into an emission signal at $z = 20$ due to the contribution of accreting BHs that efficiently heat the IGM at $z < 27$. The high-$z$ absorption feature is caused by the early coupling between the spin and kinetic temperature of the IGM induced by Pop III star formation episodes in mini-halos. Once we account for an additional radio background from early BHs, we are able to reproduce the timing and the depth of the EDGES signal only if we consider a smaller X-ray background from accreting BHs, but not the shape.
\end{abstract}




\section{Introduction}
How did the first luminous objects form? What was their impact on structure formation? These are two of the most important and still open questions regarding the evolution of our Universe. At present, we cannot rely on any direct observation of the first sources, however, we know that these were responsible for initiating the transition from a neutral intergalactic medium (IGM) to an ionized one. This crucial event in the evolution of our Universe is referred to as "cosmic reionization". According to different observations, including the quasar Gunn-Peterson trough (\citealt{Becker2001}) and the polarization anisotropy of the cosmic microwave background (CMB, \citealt{Planck2020}) the Universe became fully ionized at $z < 6$. However, the process of reionization is still largely unconstrained as we lack a complete understanding of the dominant sources (whether stars or accreting black holes), the duration of the whole process and the topology of ionized regions (see e.g. \citealt{Wise2019} for an introductory review).

Observations with the James Webb Space Telescope (JWST) will be crucial in order to better constrain the properties of high-redshift sources. However, direct optical-IR observations are not the only tool to explore the Universe at Cosmic Dawn. Before complete reionization was reached, the Universe was mostly neutral hydrogen (HI). This makes the 21cm hyperfine transition of HI and the associated global signal a very important probe of these remote cosmic epochs (we refer to \citealt{FurlanettoRev}, \citealt{Pritchard2012} and \citealt{Bera2022b} for broad reviews on the 21cm global signal from Cosmic Dawn). 

The strength and shape of the 21cm global signal depend on the ionization and thermal state of the IGM as well as on the contribution of star formation and black hole (BH) accretion to three relevant radiation backgrounds: X-ray, UV ionizing and Lyman-$\alpha$ backgrounds. In addition to those, as pointed out by \citet{Feng2018}, the presence of a radio background at high-$z$ would impact the evolution of the 21cm signal as it will change the background radio temperature against which the 21cm signal is seen. The shape and amplitude of the 21cm signal is therefore very sensitive to the nature of the first sources, probing the complex physical processes governing their formation and evolution (see e.g. \citealt{Madau1997, Ciardi2003, Mesinger2011, Visbal2012, Fialkov2013a} for some fundamental contributions). In the last decade the impact of many physical processes on the 21cm global signal has been studied; these include relative velocity between dark matter and gas (\citealt{Dalal2010,Visbal2012,munoz2021}), cosmic rays (\citealt{Bera2022a}), Ly-$\alpha$ heating (\citealt{Ghara2019}) and magnetic fields (\citealt{Katz2021}).

So far, the only claimed detection of the 21cm signal has been reported by the global signal experiment EDGES \citep{Bowman2018a}. In order to explain the large depth of the EDGES absorption feature, new physics has been taken into account; for a more detailed discussion, we refer the reader to Section \ref{sec:RadioBG}. While still controversial (e.g. \citealt{Hills2018, Bowman2018b, Sims2019, Singh2019, Bradley2019, Tauscher2020}), the signal, centered at $z \sim 17$, suggests an early start of star formation and gives a first glimpse of the amazing potential of 21cm observations that will be obtained by ongoing and forthcoming 21cm experiments, such as LOFAR (\citealt{vanHaarlem2013}, \citealt{Patil2017}, \citealt{Mertens2020}), PRIZM (\citealt{Philip2019}), NenuFAR (\citealt{Mertens2021}), HERA (\citealt{DeBoer2017}, \citealt{theheracollaboration2021hera}), REACH (\citealt{Cumner2022, deLera2022}), SKA (\citealt{Koopmans2015}) and SARAS 3 (\citealt{Singh2021}). The latter experiment is of particular relevance as it rejects the EDGES best-fitting profile with 95.3\% confidence interval. Even though so far no other detection has been claimed, the first observations of HERA already put some constraints on the temperature of the IGM at z $\sim$ 8. In particular, low X-ray scenarios seem to be unlikely, especially if a strong radio background is present \citep{theheracollaboration2021hera, Hera2022b}. However, at this stage, all the constraints on the temperature of the IGM (and therefore on the heating) are still weak and model dependent.

The purpose of this work is to compute the evolution in redshift of the 21cm global signal starting from the outputs of the recently developed semi-analytical model named \textsc{cat} (Cosmic Archaeology Tool, \citealt{Trinca2022}). \textsc{cat} was developed to investigate the early evolution of the first stars and black holes in a self-consistent way with the aim of constraining the nature of the first BH seeds and their dominant growth mode using current and future observations of the BH mass function at $z > 4$. Therefore, the model provides all the necessary input to compute the strength of the 21cm signal generated from the first stars and accreting BHs. 

Semi-analytical models (SAM) provide the simplest and fastest approach to describe the evolution of the Universe during Cosmic Dawn. They are often run on top of a merger-tree algorithm or a N-body simulation for the dark matter halo evolution in order to follow the main global properties of galaxies (gas infall and cooling, star formation etc, see \citealt{ValianteRev} for a review). By virtue of their simplicity, they allow a fast exploration of the parameter space and for this reason they have been widely used in many studies of structure formation and evolution during the reionization (see e.g. \citealt{Zhou2013, Valiante2016, Mutch2016, Mebane2020, Balu2022}). However to compute the 21cm signal and higher-order statistic (power spectrum, bi-spectrum etc.) from semi-analytical models of galaxy formation they must also account for reionization. To obtain a realistic ionization (and 21cm) map there are different approaches (see \citealt{Morales2010} and \citealt{Wise2019} for two complete reviews). These include: \textit(i) full hydro-dynamical simulations solving 3D radiative transfer equation (see \citealt{Rosdahl2018,Eide2018,Eide2020,Ocvirk2020,Garaldi2022,Kannan2022,Smith2022} for some of the most recent numerical simulation of reionization), \textit(ii) analytical computation based on the bubble model introduced by \citet{Furlanetto2004} are used to compare different reionization scenarios (\citealt{Ahn2021}) and \textit (iii) semi-numerical methods based on an excursion-set formalism to obtain an ionized distribution which is associated with a catalogue of sources through a filtering technique (\citealt{Mesinger2007}). Being much faster and computationally less expensive than numerical simulations these models are now very popular for obtaining ionization and 21cm maps \citep[e.g.][]{Mesinger2011, Fialkov2013a, Cohen2017, Qin2020, munoz2021, Jones2022}.  
In this work, reionization is computed analytically but self-consistently with the evolving sources inside \textsc{cat}. This is not a major concern of our model because we aim at predicting the 21cm global signal with a particular focus on the absorption feature during the cosmic dawn when the Universe was fully neutral. Nevertheless, \textsc{cat} allows us to describe in a self-consistent way the evolution of the first sources in the high redshift Universe (from the cosmic dawn to the epoch of reionization) including the first metal-free (or very metal-poor) stars that likely formed in low-mass halos as well as their associated black hole remnants. The modeling of these so-called light BH seeds, which dominate the BH seed population at $z \geq 15$ \citep{Trinca2022, Sassano2021, Valiante2016}, allows us to follow their subsequent growth via both gas accretion and mergers, matching the properties of SMBHs at $z \simeq 6 - 7$, as well as the evolution of the AGN luminosity function at $z \geq 4$ \citep{Trinca2022}. As far as we are aware, none of the models that follow the evolution of the first BHs has been used to compute the 21cm global signal. Throughout this work, we distinguish the impact of each class of sources (Population III stars, Population II stars and first BHs) on the evolution of the 21cm signal.

It is important to stress that \textsc{cat} models the formation of the first stars and galaxies and of their nuclear BHs under the action of radiative, chemical and mechanical feedback effects (see \citealt{Ciardi2005} for a review of feedback effects). In particular \textsc{cat} takes into account: \textit{(i)} radiative feedback both due to the emission of photons in the Lyman-Werner band, that are able to photo-dissociate molecular hydrogen and suppress star formation inside mini-halos (see \citealt{Machacek_2001,Fialkov2013b}) and the photo-heating associated to reionization, \textit{(ii)} mechanical feedback due to supernova (SN) explosions and winds powered by the energy released during BH accretion (AGN feedback) that are able to drive gas outflows from galaxies and thus suppress star formation (\citealt{Wyithe2012}) and \textit{(iii)} chemical feedback caused by SN explosions (core-collapse and pair-instability SN) that pollute the gas with metals and dust grains driving the transition from massive Population III (Pop III) stars to more enriched Population II/I (Pop II/I) stars \citep{Schneider2002, Omukai2005, Schneider2006, Chiaki2019}. 

Note that in \textsc{cat} the effect of chemical and mechanical feedback are computed according to the stellar lifetimes, thus, accounting for the typical timescales (5-30 Myr) for SN progenitors to evolve before their explosions. Thus, our model predicts a gradual chemical enrichment of the IGM that leads to a smooth transition from Pop III to Pop II/I star formation, with interesting implications for the 21cm global signal. The delay between star formation and SN explosions also has an impact on the efficiency of star formation. Because the first star forming regions are hosted by the shallow dark matter potential wells of mini-halos, we predict a model of "bursty" star formation, similar to that recently found by other semi-analytical models  \citep{Furlanetto2021}. 

The paper is organized as follows. In Section \ref{sec:21cmModel} we describe the model used for the estimation of the 21cm global signal which includes a brief description of the semi-analytical code \textsc{cat} and how we can compute all the relevant backgrounds from its outputs. Then, in Section \ref{sec:results} we illustrate the 21cm signal obtained under different assumptions and in Section \ref{sec:discussion} we discuss their features and compare these with the EDGES detection reported in \citet{Bowman2018a}. Finally we summarize the most relevant results in Section \ref{sec:conclusions}.

\section{Modeling the 21cm signal}
\label{sec:21cmModel}

The aim of this paper is to compute the 21cm background at the epoch of the formation of the first stars and BHs. Following the commonly adopted formalism introduced by \citet{FurlanettoRev}, we can express the mean 21cm global signal in terms of the differential brightness temperature (DBT) $\delta T_{\rm b}$, which represents the offset of the 21cm brightness temperature from the CMB temperature $T_\gamma$. Along a given line of sight, the DBT at an observed frequency $\nu$ can be expressed as:
\begin{equation}
    \delta {T_{\rm b}}= 26.8 \, {\rm mK} \, (1+\delta) \, x_{{\rm HI}} \, \bigg(\frac{1+z}{10}\bigg)\bigg(1-\frac{T_\gamma}{T_{\rm S}}\bigg)
    \label{eq:DBT}
\end{equation}
where $\delta$ is the matter linear overdensity, $x_{\rm HI}$ is the neutral hydrogen fraction and $T_{\rm S}$ is the spin temperature that can be written as:
\begin{equation}
    T_{\rm{S}}^{-1}=\frac{T^{-1}_\gamma+x_{\rm{c}} \, T^{-1}_{\rm{K}}+x_\alpha \, T^{-1}_{\rm{c}}}{1+x_{\rm{c}}+x_\alpha}.
    \label{eq:TS}
\end{equation}
Here $T_{\rm K}$ and $T_{\rm c}$ are the kinetic and colour temperature of the gas while $x_{\rm c}$ and $x_\alpha$ are the collisional and the Lyman-$\alpha$ coupling coefficients, which quantify the strength of the main processes that are able to alter the spin temperature of neutral hydrogen: collisions (mainly with electrons) and resonant scattering with Lyman-$\alpha$ photons through the \textit{Wouthuysen-Field} effect (see \citealt{Wouthuysen1952}). 

At the epoch of the formation of the first sources, the Universe was already expanded enough that collisions between hydrogen atoms and other species were rare making $x_{\rm c} \simeq 0$. In this case, the evolution of $T_{\rm S}$ is determined by $x_\alpha$ and the colour temperature $T_{\rm c}$, which is very close to the kinetic temperature $T_{\rm K}$. 

Following the analytical model presented by \citet{Furlanetto2006} and \citet{Pritchard2012}, we compute the Lyman-$\alpha$ coupling coefficient as:
\begin{equation}
x_\alpha = \frac{16\pi^2T_{\star}e^2f_{\alpha}}{27A_{10}T_{\gamma}m_ec} \, S_\alpha \, J_\alpha 
\label{eq:xalpha}
\end{equation}
\noindent
with numerical fits to the correction factor $S_\alpha$ and the colour temperature $T_{\rm c}$ taken from \citet{Hirata2006}. The Lyman-$\alpha$ flux, $J_\alpha$ (cm$^{-2}$ s$^{-1}$ Hz$^{-1}$ sr$^{-1}$), is computed using the input values provided by our semi-analytical model of structure formation \textsc{cat} that will be presented in Section \ref{sec:LymanBG}). The other parameters appearing in Eq. \ref{eq:xalpha} are: the Einstein coefficient for
spontaneous emission A$_{10}$, the electron mass m$_{\rm e}$, the speed of light c, the oscillatory strength of the Lyman-$\alpha$ transition f$_{\alpha}$ and the temperature equivalent to the transition energy between the two hyperfine states T$_{\star}$. In Eq. \ref{eq:xalpha} it is also highlighted the dependence of x$_{\alpha}$ on 1/T$_{\gamma}$.
Since $T_{\rm c} \simeq T_{\rm K}$, in order to compute $\delta T_{\rm b}$ from Eq. \ref{eq:DBT} we need to 
describe the thermal evolution of the neutral intergalactic gas. In an expanding Universe we can write:
\begin{equation}
    \frac{d{T_{\rm K}}}{dz}=\frac{2{T_{\rm K}}}{1+z}-\frac{2}{3}\sum_i\frac{\epsilon_i}{{k_{\rm B} n_{\rm H}}(z)(1+z)}
    \label{eq:TK}
\end{equation}
with $k_{\rm B}$ the Boltzmann constant, $n_{\rm H}$ the hydrogen gas number density and $\epsilon_i$ the heating rate per unit volume of the $i$-th process. 

Throughout this work we consider X-rays as the only source responsible for gas heating, following other semi-analytical models of the 21 cm global signal \citep[e.g.][]{Mesinger2011,Cohen2017,Chatterjee2019}. As pointed out by \citet{Furlanetto2006}, Lyman-$\alpha$ photons and shocks can provide additional contributions to gas heating. The first has been shown to be effective in models with low X-ray heating (\citealt{Chuzhoy2007,Ciardi2007,Ciardi2010,Reis2021,Mittal2021}), however in many scenarios it should be subdominant compared to X-ray heating. While the second is still higly debated; according to some hydrodynamical simulations \citep[e.g.][]{Kuhlen2006} shocks do not strongly modify the thermal history of the gas, but other studies predict a stronger impact \citep[e.g.][]{Furlanetto2004,Gnedin2004,Ma2021}. Another contribution that will not be considered in this work, is the heating by cosmic rays which, as pointed out by \citet{Bera2022a}, might be significant at high redshift. A presentation of the adopted model to describe X-ray heating will be given in Section \ref{sec:XBG}.

The strength of the 21cm signal described by Eq. (\ref{eq:DBT}) is determined by the amount of neutral hydrogen $x_{\rm HI}$, the radio background, the cosmological parameters and the spin temperature evolution. During cosmic reionization, the neutral hydrogen fraction decreases with time
and assuming that the IGM is made only by hydrogen, we can write $x_{\rm HI} = 1 - \bar{x}_i$ where $\bar{x}_i$ is the global ionized hydrogen fraction. The spin temperature evolution is instead determined by the evolution of the gas kinetic temperature, $T_{\rm K}$ (since this is very close to $T_{\rm c}$) and by the Lyman-$\alpha$ coupling coefficient, $x_\alpha$. 

The computation of $\bar{x}_i$, $T_{\rm K}$ and $x_\alpha$, and their evolution in redshift, requires a quantification of the time-dependent radiation field in three different wavebands: \textit{(i)} UV ionizing band, which determine the reionization history, \textit{(ii)} X-ray band, which is able to heat up the gas and \textit{(iii)} Lyman-$\alpha$ band, which is responsible for the coupling between $T_{\rm S}$ and $T_{\rm c}$. Once the first sources appear in the Universe, they emit photons at all these relevant wavelengths, and their effects are encoded in the shape of the 21 cm signal. Hence, a model for structure formation is required, as described in the following section.

\subsection{The Cosmic Archaeology Tool (\textsc{cat})}
\label{sec:CAT} 

The semi-analytical code adopted throughout this work is the Cosmic Archaeology Tool (\textsc{cat}, \citealt{Trinca2022}). 
A thorough description of the model can be found in the original paper, where it has been used to investigate the  evolution of galaxies and their nuclear black holes at $z \geq 4$. \textsc{cat} describes the formation of the first stars and black hole seeds in a self-consistent way, following the feedback-regulated co-evolution of nuclear BHs and their host galaxies from cosmic dawn down to the post-reionization era (\citealt{Trinca2022}). As such, \textsc{cat} provides us with the information required to estimate the strength of the 21cm global signal generated from the first stars and BHs. 

\textsc{cat} uses the model \textsc{galform} to describe the redshift evolution of dark matter halos (\citealt{Cole2000,Parkinson2008}), while the description of star formation, chemical evolution, BH seeding and growth is imported from the semi-analytical model \textsc{gqd} \citep{Valiante2011,Valiante2016,Sassano2021}. \textsc{cat} simulations have been run assuming a $\Lambda$CDM model and taking the cosmological parameters according to \citet{Planck2020}: $\Omega_\Lambda=0.685$, $\Omega_m=0.315$, $\Omega_b=0.05$, h=0.674 and $\sigma_8=0.81$.

\subsubsection{Halo merger trees}

\textsc{galform} is a semi-analytical model of galaxy formation that is able to reconstruct the hierarchical merger history (or merger tree) of a given dark matter (DM) halo. This Monte Carlo algorithm, based on the Extended Press-Schechter theory, was originally developed by \citet{Cole2000} and then improved by \citet{Parkinson2008} who perturbed the basic function that drives the algorithm obtaining first-order corrections. \textsc{galform} starts with a DM halo at redshift $z_0$ of a given mass and follows its evolution back in time reconstructing its progenitors. In \textsc{cat}, the \textsc{galform} algorithm has been used to generate merger trees for DM halos with masses $10^{9}\,M_\odot \leq M \leq10^{14} \,M_\odot$ at $z_0 = 4$. This mass range has been divided into 11 logarithmically spaced bins with size 0.5 and for each bin a final halo of mass equal to the central bin value has been considered as a starting point for the code to simulate 10 independent halo merger trees; the total merger tree sample accounts thus for 110 merger trees. This choice constitutes a representative sample of the halo population at $z \geq 4$ (see \citealt{Trinca2022}). The resulting redshift dependent mass distributions of each mass bin are weighted according to the number density of DM halos at redshift $z_0 = 4$, as given by the Sheth and Tormen mass function \citep{Sheth2002}. This model includes some free parameters which are tuned so that the resulting merger histories are in agreement with the N-body Millennium simulation \citep{Springel2005}. Differently from \citet{Trinca2022}, in this work we extended the merger trees to higher redshifts as the 21 cm absorption feature is expected to be seen at the cosmic dawn.

In our setup, we adopt a resolution mass (the minimum mass of a collapsed halo identified in the merger trees) that depends on redshift and corresponds to a virial temperature of $T_{\rm vir} = 1200 \, {\rm K}$, where \citep{Bromm2013}:
\begin{equation}
M_{\rm halo} = 10^6 {\rm M_\odot} \,  \biggl( \frac{T_{\rm vir}}{2 \times 10^3 \, \rm{K}}\biggl)^{3/2} \, \biggl( \frac{1+z}{20} \biggl)^{-3/2}.
\end{equation}
\\
\noindent 
With this choice, we are able to describe the formation history of \textit{mini-halos}, where molecular hydrogen gas cooling allows the formation of the first stars at $z \sim 20 - 30$. For the purposes of the present work, since the 21 cm absorption feature is expected to originate at cosmic dawn, we extended the merger trees produced with GALFORM to higher redshifts with respect to \citet{Trinca2022}. The maximum redshift for the onset of the calculation is $z_{\rm max}$ = 40, and 800 time steps logarithmically spaced between $z_{\rm max}$ and $z = 4$ are adopted. This makes time steps larger for smaller redshifts ($\sim 5$ \, Myr at $z \sim 4$, $\sim 0.2$\, Myr at $z \sim 40$). The main consequences of the above time resolution are that merger events can involve more than two DM halos, allowing for multiple mergers and that we can resolve supernovae feedback at all redshifts. 

\subsubsection{Gas accretion and infall}
In this work we decided to follow the evolution of the first mini-halos starting from very high redshift, $z_{\rm max} = 40$. In order to properly characterize the first episodes of star formation, we included additional physical processes which are not usually resolved/included in SAMs. These processes play an important role in regulating gas cooling, especially in the first collapsed halos.
During the initial collapse of a dark matter halo, the kinetic energy of the infalling gaseous component is transformed into thermal energy through shocks and dissipation. Therefore, it is usually assumed that the gas is heated up to the virial temperature of the host halo. Subsequent episodes of mass growth will result in additional heating for the gas in the newly formed galaxy, possibly leading to a delay in the process of gas cooling and, consequently, in the onset of star formation.
Following the analysis of dynamical heating originally presented by \citet{wang2008}, we assign to each dark matter halo a gas heating rate: 

\begin{eqnarray}
\Gamma&=&-2.95\times10^{-8}\left({M_{\rm halo}\over
M_\odot}\right)^{-{1\over3}}{d(M/M_\odot)\over
dz}\cr &&\times[\Omega_{m}(1+z)^7+\Omega_{\Lambda}(1+z)^4]^{1/2} \cr &&\times
\left({\Omega_{m}\over0.3}\right)^{{1\over3}}\left({\Delta_c\over178}\right)^{{1\over3}}\left({\mu\over0.59}\right)
{\rm eV} \, {\rm Gyr}^{-1} \ , \label{2.5}
\end{eqnarray}
where $\Delta_c$ is the characteristic overdensity at the virial radius, $\mu$ is the mean molecular weight and $dM/dz$ is the halo mass accretion rate, which in our model is entirely determined by the halo merger tree. 
At each time step, if the heating rate associated with halo mass growth is higher than the gas cooling rate estimated inside the galaxy \citep[for a detailed description see Appendix A in][]{Valiante2016} we assume that the gas is not able to cool down efficiently and, as a result, star formation is inhibited.

In addition, large-scale streaming velocities of the baryonic components relative to dark matter, which results from the coupling between baryons and radiation prior to the epoch of recombination, may lead to a delay in the appearance of the first stars, suppressing the star formation in the first low-mass halos.
Relying on the analysis done by \citet{schauer2021}, performed using state-of-the-art hydrodynamical simulations, we assume that the minimum halo mass for the onset of star formation due to the impact of streaming velocities is
\begin{equation}
    \log_{10}M_\mathrm{min} = \log_{10} M_0 + s\times \frac{v_\mathrm{bc}}{\sigma_\mathrm{rms}} ,  \label{eq:fit3}
\end{equation}
with
\begin{equation}
\log_{10} M_0 = 5.562 , \,\,\, s = 0.614 
\end{equation}
and where we assumed the streaming velocity of the baryonic component to be $v_\mathrm{bc} = 1 \sigma_\mathrm{rms}$.

\subsubsection{Star formation and feedback}
Along a merger tree, each progenitor galaxy can form stars according to the available gas mass, $M_{\rm gas}$. The corresponding star formation rate (SFR) is given by \citep{Valiante2016, Sassano2021}:
\begin{equation}
    {\rm{SFR}}=\frac{f_{\rm cool} \,  f_\star \, M_{\rm gas}}{\tau_{\rm{dyn}}},
    \label{eq:SFR}
\end{equation}
where $\tau_{\rm dyn}$ is the halo dynamical time, $f_\star$ is the star formation efficiency (one of
the free parameters of the model) and $f_{\rm cool}$ quantifies the cooling efficiency. A number of
feedback processes have a key role in the evolution of star formation as they regulate the quantities entering in Eq. (\ref{eq:SFR}). \textsc{cat} accounts for them through the following prescriptions:
\begin{itemize}
    \item\textit{Pop III star formation}\\
     In order to properly follow stellar evolution in the first mini-halos we improved the prescription for Pop III star formation (Trinca et al. in prep) based on the result of recent hydrodynamical simulations by \cite{Chon2021}. For pristine halos, we set a minimum mass of cold gas formed in a dynamical timescale of $M_{\rm cold} = f_{\rm cool} \,  M_{\rm gas}/ \tau_{\rm{dyn}} = 10^3 \, M_{\odot}$, above which the gas cloud is assumed to be gravitationally unstable and it proceeds to collapse. Below this threshold value the halo will not experience star formation but accumulates cold gas for subsequent star formation. Moreover, if Pop III star formation occurs in a mini-halo, an enhanced efficiency $f_{\star} = f_{\star, \rm{Pop III}} = 0.15$ is assumed, so that the minimum total stellar mass formed in a single episode will be $M_{\rm min, PopIII} = 150 M_{\odot}$, as suggested by \citet{Chon2021}. \\

    \item\textit{Radiative feedback}\\
    The external UV radiation illuminating the galaxy regulates $f_{\rm cool}$ and $f_\star$ through two main processes: photo-dissociation of H$_2$ and photo-heating. For a gas of primordial composition, in mini-halos ($1200 \,{\rm K} \leq$ T$_{\rm vir}< 10^4 \,{\rm K}$) cooling occurs through molecular hydrogen, while in more massive atomic cooling halos (T$_{\rm vir} \geq 10^4 \,{\rm K}$) atomic cooling becomes efficient. The efficiency of molecular cooling depends on the amount of H$_2$. This molecule can be easily photo-dissociated by Lyman-Werner (LW) photons, suppressing star formation and this effect is quantified by the parameter $f_{\rm cool}$. In particular, we set $f_{\rm cool} = 1$ in atomic cooling halos while lower values are taken in mini-halos depending on $T_{\rm vir}$, redshift, gas metallicity and Lyman-Werner flux intensity, J$_{\rm LW}$ (see \citealt{Valiante2016}, \citealt{deBennassuti2017} and \citealt{Sassano2021} for more details). However, we do not account for self-shielding effects that would decrease the strength of the J$_{\rm LW}$, reducing the effect of photo-dissociating (or LW) feedback \citep[e.g.][]{munoz2021}. 
    
    Unlike photo-dissociation that only affects star formation inside mini-halos, the increased temperature due to hydrogen photo-ionization can inhibit star formation even inside atomic cooling halos (\citealt{Hui1997}, \citealt{Graziani2015}). To account for this effect, we take $f_\star=0$ when $T_{\rm vir}$ is below the IGM temperature $T_{\rm IGM}$. This is computed as $T_{\rm IGM}=\bar{x}_i(z) \, T_{\rm reio}+[1 - \bar{x}_i(z)]T_{\rm gas}$ where $T_{\rm reio} = 2 \times 10^4$\,K is the post-reionization temperature, and $T_{\rm gas} = 170 \, {\rm K} [(1+z)/100]^2$.\\
    
    \item\textit{Mechanical feedback}\\
    Mechanical feedback affects the amount of gas available for star formation ($M _{\rm gas}$). The two relevant processes that drive gas outflows from galaxies are SN explosions and the winds powered by the energy released during BH accretion (AGN feedback). The corresponding gas outflow rates are described as \citep{Valiante2011, Valiante2016}:
    \begin{eqnarray}
    \dot{M}{\rm{_{ej,SN}}}=\frac{2 \, E_{\rm SN} \epsilon_{\rm{{w,SN}}}R_{\rm{{SN}}}(t)}{\nu_e^2}\\
    \dot{M}{\rm{_{ej,AGN}}}=2 \,\epsilon_{\rm{{w,AGN}}} \, \epsilon_{\rm{r}} \, \dot{M}_{\rm{{accr}}}\bigg(\frac{c}{\nu_{\rm{e}}}\bigg)^2,
    \end{eqnarray}
    where R$_{\rm {SN}}(t)$ is the SN explosion rate and E$_{\rm {SN}}$ the average explosion energy per SN ($2.7\times10^{52}$ erg for Pop III stars, and $1.2\times10^{51}$ erg for Pop II/I stars). $\dot{M}_{\rm accr}$ is the gas accretion rate on the nuclear BH, $\epsilon_{\rm r}$ the BH radiative efficiency (see section \ref{subsec:BH}) and $\nu_{\rm e}$ is the escape velocity of the gas. The strength of mechanical feedback depends on two free parameters of the model: the SN- and AGN-driven wind efficiencies ($\epsilon_{\rm w,SN}$ and $\epsilon_{\rm w,AGN}$, respectively).\\
    
    \item\textit{Chemical feedback}\\
    Once formed, the stars are distributed in mass according to a Larson Initial Mass Function (IMF, \citealt{Larson1998}):
    \begin{equation}
        \phi(m_\star)\propto m_\star^{\alpha-1}\exp(-m_{\rm ch}/m_\star),
        \label{eq:LarsonIMF}
    \end{equation}
    where $\alpha = -1.35$ and $m _{\rm ch}$ is a characteristic mass and it is taken to be $20 \, M_\odot$ for Pop III stars (which are assumed to form with masses is the range $10 \, M_\odot \leq m_\star \leq 300 \,M_\odot$, \citealt{deBennassuti2017}) and $0.35 \, M_\odot$ for Pop II/I stars, which have masses in the range $0.1 \, M_\odot \leq m_\star \leq 100M_\odot$. 
   
    The critical metallicity that defines the transition from Pop III to Pop II/I stars is set to be $Z_{\rm cr}=10^{-3.8} \, Z_\odot$, motivated by the physical processes occurring in low-metallicity star forming clouds \citep{Omukai2005, Schneider2006, Schneider2012} and by stellar archaeology studies \citep{deBennassuti2014, deBennassuti2017}, although more recent studies seem to suggest a gradual transition from a top-heavy to a bottom-heavy stellar IMF \citep{Chon2021, Chon2022}. 
    
    Evolving stars progressively enrich the IGM with metals and dust, with mass- and metallicity dependent yields that are taken from \citet{vandenHoek1997} and \citet{Zhukovska2008} for AGB stars (1 - 8 \, $M_\odot$), \citet{Woosley1995} for core-collapse SNe and \citet{Heger2002} for pair-instability SNe. Dust yields are taken from \citet{Bianchi2007, Zhukovska2008}. For a more detailed description of the chemical evolution model with dust we refer the reader to \citet{Valiante2014} and  \citet{deBennassuti2014}.
\end{itemize}

\subsubsection{Black hole formation and growth}
\label{subsec:BH}
Supermassive BHs are considered to form via both gas accretion and mergers starting from less massive progenitors, referred to as \textit{seeds}. These are commonly divided by mass into different populations: light seeds ($\sim 100 \, M_\odot$) coming from Pop III stellar remnants (\citealt{Haiman2001} \citealt{Madau2001}), medium-weight seeds ($\sim 10^3 \, M_\odot$) formed by run-away collisions of stars in dense clusters \citep{Omukai2008, Devecchi2010, Sassano2021}, and heavy seeds ($\sim 10^4-10^5 \, M_\odot$) formed after a direct collapse of a giant molecular cloud mediated by the formation of a super-massive star (\citealt{Bromm2003}, \citealt{Johnson2012}). After the formation of BH seeds, we need to consider their growth via gas accretion and coalescence. Following \citet{Valiante2011, Valiante2016, Sassano2021, Trinca2022} we assume that a merger event between two BHs can occur only during major galaxy mergers. Therefore, when the mass ratio of their interacting host DM halos is $\mu > 1/10$ (considering two DM halos with $M_1 > M_2$, $\mu \equiv M_2/M_1$) we assume the nuclear BHs to merge within the typical timescale of the simulation $\Delta t \sim 1 \rm \, Myr$. 

BH accretion is described by the Bondi-Hoyle-Lyttleton (BHL) accretion rate, given by (\citealt{Bondi1952}):
\begin{equation}
    \dot{M}_{\rm{{BHL}}}=\alpha \, \frac{4\pi G M^2_{\rm BH} \, \rho_{\rm gas}(r_{\rm{A}})}{{c}^3_{\rm{s}}}
    \label{eq:BHL}
\end{equation}
\noindent
where $c_s$ is the sound speed and $\rho_{\rm gas}(r_{\rm A})$ is the gas density evaluated at the Bondi radius, $r_{\rm A} = 2 G M_{\rm BH}/c^2_{\rm s}$ (i.e. the radius of gravitational influence of the BH). The parameter $\alpha$ is a free parameter of the model which does not appear in the original BHL model but it is usually introduced to account for the enhanced gas density in the inner regions around the central BH \citep{Valiante2016}\footnote{A value of $\alpha = 90$ is assumed for black holes hosted in dark matter halos with $M_{\rm BH}/M_{\rm halo} \leq 1/2000$. This condition applies to systems where the Bondi radius of the BH becomes smaller than $0.1$ times the core radius of the gas density profile, which is assumed to be isothermal with a flat core (see \citealt{Trinca2022}).}.  For the present study, we assume that BH accretion does not exceed the Eddington rate , so that $\dot{M}_{\rm accr}=\min(\dot{M}_{\rm BHL},\dot{M}_{\rm Edd}$) with $M_{\rm Edd}$ computed from the standard definition of the Eddington luminosity $L_{\rm Edd}$ as $\dot{M}_{\rm Edd}=L_{\rm Edd}/\epsilon_{\rm r} c^2$ and taking $\epsilon_{\rm r} = 0.1$.

The physical processes presented above describe star formation and BH evolution with 4 free parameters: $f_\star$, $\epsilon_{w,SN}$, $\epsilon_{w,AGN}$ and $\alpha$. We select these parameters to be consistent with the reference model presented in \citet{Trinca2022}, that provides predictions for the galaxy and AGN populations that are in good agreement with observational data and a reionization history consistent with available observational constraints (Trinca et al. in prep.). Hence, in the following we will assume $f_\star$= 0.05, $\epsilon_{\rm w,SN}=1.6 \times 10^{-3}$, $\epsilon_{\rm w,AGN}=2.5 \times 10^{-3}$ and $\alpha = 90$. By running \textsc{cat} simulations with this set of parameters, we are able to predict the properties of each galaxy at each redshift for all the merger tree simulations that are needed to estimate the radiation fields necessary for the computation of the 21cm global signal, as shown below.

\subsection{Ionizing UV background}
\label{sec:UVBG}

The amount of UV photons produced by the first stars and BHs is crucial to compute the reionization history and it has a direct impact on the evolution of the 21cm global signal. Following \citet{Barkana2001}, we estimate the ionized hydrogen fraction as:
\begin{eqnarray}
    \bar{x}_{\rm{i}}(z_{\rm{{obs}}})=\frac{f_{\rm{esc}}}{\bar{n}^0_{\rm{H}}}\int^\infty_{z_{\rm{{obs}}}}dz \, \bigg|\frac{dt}{dz}\bigg| \, \dot{n}_{\rm{{ion}}} \, \, e^{F(z_{\rm{{obs}}},z)},
    \label{eq:xi}\\
    F(z_{\rm{{obs}}},z)=-\alpha_{\rm{B}} \, \bar{n}^0_{\rm{H}}\int_{z_{\rm{{obs}}}}^z dz' \, \bigg|\frac{dt}{dz'}\bigg|\, C(z')(1+z')^3,
    \label{eq:F}
\end{eqnarray}
where $f_{\rm esc}$ is the escape fraction of ionizing photons and $\dot{n}_{\rm ion}(z)$ is the ionizing photon rate density (number of photons cm$^{-3}$ s$^{-1}$). We compute the latter quantity from the luminosities of the stellar populations and accreting BHs predicted by \textsc{cat}. The term $F(z_{\rm obs},z)$ accounts for the recombination rate of ionized hydrogen. It depends on the clumping factor $C$, which quantifies the patchiness of the IGM, on the present-day hydrogen number density $\bar{n}^0_{\rm H}$ and on the case-B recombination factor $\alpha_{\rm B} = 2.6\times10^{-13}$cm$^3$\, s$^{-1}$ (valid for hydrogen at $T = 10^4$\,K, \citealt{Barkana2001}). The clumping factor and $f_{\rm esc}$ are free parameters, that we tune in order to obtain ionization histories consistent with the latest \citet{Planck2020} measurements and with the ionizing photon emissivities constrained from observations at $z < 6$ \citep{Bolton2007, Kuhlen2012, Becker2013, Becker2021}.

A thorough description of the reionization scenarios predicted by \textsc{cat} will be presented in Trinca et al. (in prep). In the following, we adopt a redshift-dependent escape fraction $f_{\rm esc}(z) = 0.02 \times [(1+z)/5)]^{3/2}$, in agreement with what has been done in other recent studies \citep[e.g][]{Mutch2016}.

\begin{figure*} 
    \includegraphics[width=\columnwidth]{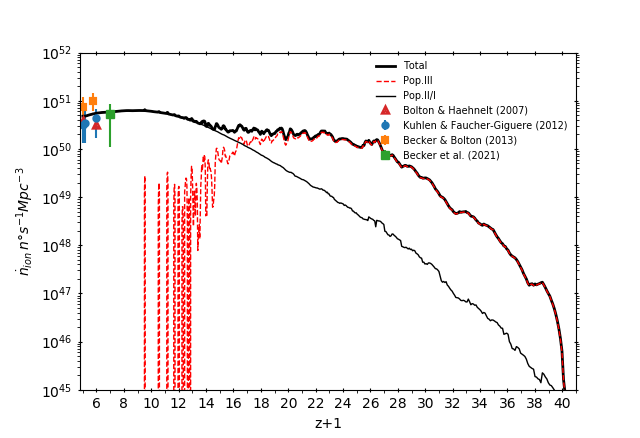}
    \includegraphics[width=\columnwidth]{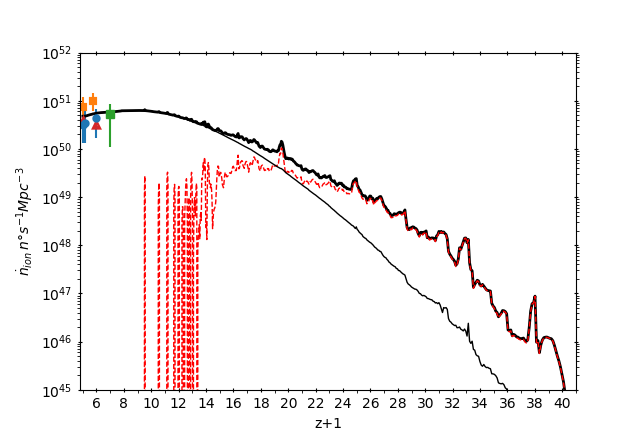}
    \caption{Ionizing photon rate density (number of photons Mpc$^{-3}$ s$^{-1}$) as a function of redshift predicted by \textsc{cat} for the fiducial model, considering all the galaxies (left panel) and without the contribution of galaxies forming in mini-halos. For each panel we highlight the main contributions: Pop II/I (thin black line) and Pop III (dashed red line) stars (both multiplied by the escape fraction of ionizing photons $f_{\rm esc}$ as described in the text). The model predictions are compared with different observations, as labeled in the figure: \citet{Bolton2007}, \citet{Kuhlen2012}, \citet{Becker2013} and \citet{Becker2021}.} 
    \label{fig:UvIonRun5}
\end{figure*}

At each redshift, \textsc{cat} is able to compute for each galaxy its SFR ($M_\odot$/yr), stellar metallicity (this is crucial to distinguis between Pop III and Pop II/I stars), nuclear BH mass and accretion rate. The stellar contribution to the ionizing (E > 13.6 eV) photon rate can be recovered using time- and metallicity-dependent UV luminosities from \citet{Bruzual2003} for Pop II/I stars and from \citet{Schaerer2002} for Pop III stars (see also Fig. 3 in \citealt{deBennassuti2017}). 
Following \citet{Valiante2016} and \citet{Trinca2022}, the ionizing UV photon rate produced by each accreting BH depends on the Eddington ratio and on the BH mass. We then account for obscuration in the UV band, assuming an obscured fraction of \citep{Merloni2013}:
\begin{equation}
    f_{\rm obs}= A + \frac{1}{\pi}\arctan\bigg(\frac{l_0-L_{\rm X,2-10\,keV}}{\sigma_{\rm X}}\bigg),
    \label{eq:Fobs}
\end{equation}
where $L_{\rm X,2-10\,keV}$ is the logarithmic X-ray AGN luminosity in the 2 - 10 keV band and the best-fit parameters are: $A = 0.56$, $l_0=43.89$ and $\sigma_{\rm X}=0.46$. The procedure we adopt to compute $L_{\rm X,2-10\,keV}$ is described in Section \ref{sec:XBG}. In our model, we assume that only a fraction equal to $1-f_{\rm obs}$ of all AGNs is unobscured and contributes to the ionizing photon emissivity.

The resulting $\dot{n}_{\rm ion}$ multiplied by the above $f_{\rm esc}$ is shown in the left panel of Fig. \ref{fig:UvIonRun5}, with the separate contributions coming from Pop III and Pop II/I stars (these are the two dominant contributions). We compare our models with some observational constraints at $z < 6$ as indicated in the Figure. Overall, the dominant contribution is provided by Pop II/I emission (thin black line), with Pop III stars (dashed red line) dominating during the first phase of the evolution, at $z + 1 > 20$ and disappearing at $z + 1 < 10$ as a consequence of the increased level of metal enrichment in the galaxies and the IGM. The Pop III contribution to the ionizing photon rate has a less smooth evolution compared to the Pop II/I especially in the redshift interval $z + 1 \sim 10 - 16$. 
The right panel of Fig. \ref{fig:UvIonRun5} shows the same quantities but eliminating the contributions of galaxies 
hosted by mini-halos (systems with $T_{\rm vir} < 10^4$ K). Both Pop III and Pop II contribution at $z + 1 < 15$ are not affected with the Pop II contribution largely dominating over the Pop III one. However, at high-$z$ both contributions are weaker. The impact on the ionizing emission from Pop II stars is only mildly reduced, while the Pop III one suffers of a much stronger suppression ($\sim$ 1 order of magnitude at $z + 1 > 22$). In addition, this latter contribution is more "bursty". This evolution reflects the fact that at high-$z$ a large fraction of Pop III (and a smaller fraction of Pop II) star formation occurs in mini-halos. Conversely, at lower redshift, most of the star formation occurs in metal enriched atomic cooling halos. At $z + 1 < 15$ the emission of ionizing photons is halted as a consequence of the effect of the LW background that suppress molecular cooling in these less massive systems.

Overall, our model predicts a production of ionizing photons at $z + 1  = 5 - 7$ consistent with observations reported by \citet{Bolton2007}, \citet{Kuhlen2012} \citet{Becker2013} and \citet{Becker2021}. In addition, depending on the adopted clumping factor\footnote{In Trinca et al. (in prep) we explore different reionization histories depending on the adopted evolution of the clumping factor $C(z)$. The values quoted above refer to a model obtained using the maximum value between a constant clumping factor of 3 and the analytic form presented in \citet{Iliev2007} ($z_{\rm reio} = 4.67$, $\tau_e = 0.050$), and a model obtained using a constant clumping factor of 3 ($z_{\rm reio} = 6$, $\tau_e = 0.065$).}, we obtain a total Thomson scattering optical depth of $\tau_e = 0.050 - 0.065$, consistent with the latest measurements of \citet{Planck2020}, and a complete reionization redshift of $z_{\rm reio} = 4.67 - 6$. All the models predict the same high-redshift reionization history with an optical depth in the redshift range $15 \leq z \leq 30$ of $\tau_e (15 - 30) = 0.050 - 0.065$. We refer to Trinca et al. (in prep) for a more in-depth discussion of the reionization histories predicted by \textsc{cat}.

\subsection{Lyman-$\alpha$ background} 
\label{sec:LymanBG}

The Lyman-$\alpha$ background impacts the evolution of the 21cm global signal as it determines the coupling between the kinetic and spin temperature through the coupling coefficient $x_\alpha$, which in turn depends on the Lyman-$\alpha$ flux $J_\alpha$ and on its evolution in time. 

Following \citet{Dayal2008}, we simply assume the Lyman-$\alpha$ photon rate to be set by the amount of UV ionizing photons produced by stars and accreting BHs which do not escape from the galaxies where they are generated, hence $\dot{n}_{\alpha} = \dot{n}_{\rm{ion}} \, (1-f_{\rm{esc}})$.
Thus, the Lyman-$\alpha$ flux is computed as:
\begin{equation}
    J_\alpha(z_{\rm{obs}})\simeq\frac{c}{4\pi}\int_{z_{\rm{obs}}}^\infty dz \, \bigg|\frac{dt}{dz}\bigg| \, \frac{(1+z)^2}{H(z)} \, \frac{\dot{n}_{\rm{ion}}(1-f_{\rm{esc}})}{\Delta\nu} \, f_\alpha,
    \label{eq:Jalpha}
\end{equation}
with $\Delta\nu=\nu_{\rm LL}-\nu_\alpha$ the frequency interval between the Lyman limit and the Lyman-$\alpha$ line and $f_\alpha$ the fraction of Lyman-$\alpha$ photons that escape the galaxy without being destroyed by dust, that we assume to be equal to 1\footnote{Here we are mostly interested in the 21cm global signal at cosmic dawn, hence we do not expect significant dust enrichment in the interstellar medium of galaxies at $z > 15$.}. The main advantage of this formalism is that we do not need any new quantity to compute $J_\alpha$ since we use the ionizing UV photon rate already discussed in the previous section.

\begin{figure}
    \includegraphics[width=\columnwidth]{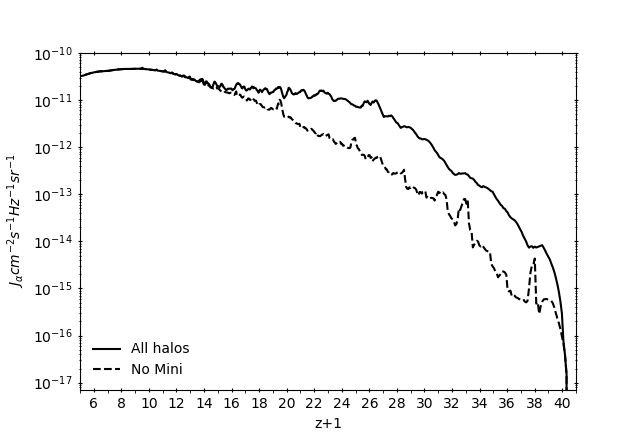}
    \caption{Redshift evolution of the Lyman-$\alpha$ flux $J_\alpha$ [cm$^{-2}$\, s$^{-1}$\, Hz$^{-1}$\,sr$^{-1}$] computed from \textsc{cat} (see text). The solid and dashed lines show the results from sources in all halos and without mini-halos, respectively.}
    \label{fig:JalphaRun5}
\end{figure}

Fig. \ref{fig:JalphaRun5} shows the results including all stars and accreting BHs predicted by \textsc{cat} (solid line), and without the contribution of mini-halos (dashed line). The evolution of the Lyman-$\alpha$ flux shows the same trend of the UV ionizing emission (see Fig. \ref{fig:UvIonRun5}). At high redshift, sources hosted in mini-halos strongly contribute to the Lyman-$\alpha$ background while for $z + 1 \leq 15 $ the evolution of ${\rm J_ \alpha}$ is determined by sources in atomic cooling halos. Thus, the contribution of mini-halos is relevant in determining a coupling between $T_{\rm S}$ and $T_{\rm K}$ at early times, while at later times it becomes negligible.

\begin{figure*}
    \includegraphics[width=\columnwidth]{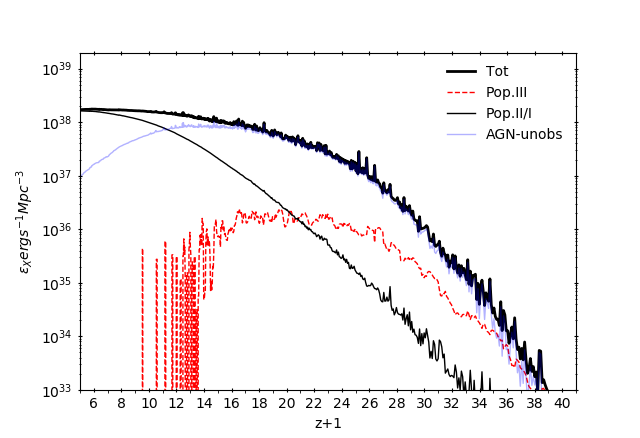}
    \includegraphics[width=\columnwidth]{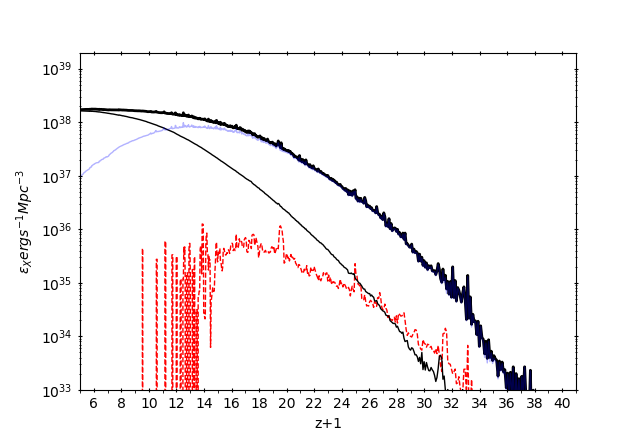}
    \caption{Redshift evolution of the soft X-ray emissivity in the (units of erg s$^{-1}$ Mpc$^{-3}$, black line) with all the galaxies predicted by \textsc{cat} (left panel) and without the contribution of systems hosted in mini-halos (right panel). In both panels, we show the separate contributions from Pop II/I stars, Pop III stars and unobscured AGNs adopting the same line coding as in Fig. \ref{fig:UvIonRun5}.}
    \label{fig:XheatRun5}
\end{figure*}

\subsection{X-ray background}
\label{sec:XBG}

The X-ray background is responsible for gas heating and determines the evolution of the kinetic temperature of the gas, as shown by Eq. (\ref{eq:TK}). In particular, we need to estimate the X-ray emissivity $\epsilon_{\rm X}$ (erg s$^{-1}$ Mpc$^{-3}$) produced by X-ray binaries and accreting BHs.

Following \citet{Grimm2003}, the first contribution is modeled according to the locally observed correlation between the star formation rate and the bolometric X-ray luminosity, $L_{\rm X}$ (erg s$^{-1}$):
\begin{equation}
    L_{\rm X}=3.4\times10^{40} \, f_{\rm X} \, \frac{{\rm SFR}}{M_\odot {\rm yr}^{-1}}.
    \label{eq:LX}
\end{equation}
This relation is based on CHANDRA and ASCA observations of nearby star-forming galaxies. In particular, the value of the constant correctly matches the proportionality between the X-ray luminosity of a galaxy arising from high-mass X-ray binaries (HMXBs) and the SFR for high values of SFR. The contribution of low-mass X-ray binaries (LMXBs) is not considered as it has been found to be subdominant with respect to the one coming from HMXBs at $z \geq 2.5$ (\citealt{Fragos2013,Madau2017}). Our ignorance of the accuracy of this relation at high-redshift (in particular the proportionality constant) is factorized in the parameter $f_{\rm X}$\footnote{This parameter likely will evolve with redshift and/or metallicity but for simplicity we assume it to be a constant. The dependence on metallicity though has not a very strong impact as found in \citet{Kaur2022}.}.
We can thus express $\epsilon_{\rm X}$ as \citep{Chatterjee2019}:
\begin{equation}
    \epsilon_{\rm X}=3.4\times10^{40} \, {\rm erg} \, {\rm s}^{-1} \, {\rm Mpc}^{-3} \, f_{\rm X} \, f_{\rm{X,h}} \, \frac{\dot{\rho}_\star}{M_\odot \, {\rm yr}^{-1} \, {\rm Mpc}^{-3}},  
    \label{eq:epsX}
\end{equation}
where $f_{\rm X,h} = 0.2$ is the fraction of the X-rays that contribute to heating \citep{Shull1985}. 

In this work we assume $f_X=1$. This choice is consistent with the recent results of the HERA collaboration \citet{theheracollaboration2021hera}, which imply that galaxies at higher redshift were more efficient at producing X-rays than local ones, constraining the ratio between $L_{\rm X}$ and SFR between 1.6 $\times 10^{40}$ and 8 $\times 10^{41}$. However, it is worth noting that HERA observations are taken at z = 8 and z = 10 and do not provide constraints at earlier times.

Finally, we need to account for the fact that X-ray photons will not be equally efficient at heating the IGM. The higher the energy of the photon, the higher its mean free path $\lambda_{\rm X}$, and lower the probability of interacting with the IGM. As shown by \citet{Fialkov2014} and more recently by \citet{Madau2017} a photon will not interact if $\lambda_{\rm X} > c/H(z)$, yielding an upper limit on the energy of X-ray photons that can be absorbed by the IGM of:
\begin{equation}
    E_{\rm lim}=1.7\bigg(\frac{1+z}{10}\bigg)^{0.47}{\rm keV}
    \label{eq:Elim}
\end{equation}
which gives $E_{\rm lim} \in $ [1.23, 3.29] keV in the redshift interval of our simulation.
Following \citet{munoz2021}, we adopt a power-law spectral energy distribution (SED) for HMXBs with a spectral index $\alpha_{\rm X} = -1$ and a low-energy cut-off $\rm{E}_0 = 0.5$. This SED is normalized to the bolometric X-ray emissivity computed from Eq. (\ref{eq:epsX}). This choice is broadly consistent with emission spectra of HMXBs in low-metallicity environments \citep[e.g][]{Fragos2013, Madau2017, Das2017, Qin2020}. We highlight that we are considering a simple SED that we apply to Pop III and Pop II/I stars. This is a crude approximation, and we defer to a future work a more in-depth analysis of the X-ray SED.
The X-ray emissivity that enters in Eq. (\ref{eq:TK}) is then $\epsilon_{\rm X, E<E_{lim}} = \int_{E_0}^{ E_{\rm lim}}\epsilon_{\rm X}(E)dE$. 

The X-ray contribution from BHs is calculated by applying a correction factor $K_{\rm X}$ to the bolometric luminosity of the AGN so that $L_{\rm X, 2 - 10\, keV} = L_{\rm bol}/K_{\rm X}$. We take the correction factor from \citet{Duras2020}:
\begin{equation}
    K_{\rm X}= a \bigg[1+\bigg(\frac{\log(L_{\rm bol}/L_\odot)}{b}\bigg)^c\bigg],
    \label{eq:KX}
\end{equation}
where $a = 10.96, b = 11.93$, and $c = 17.79$. This relation is valid for both type 1 and type 2 AGNs (the sampled analysed included $\sim$1000 type 1 and type 2 AGNs). The hard X-ray spectrum of AGNs is important to determine the AGN obscuration fraction in the UV band (see Eq. (\ref{eq:Fobs})). As stressed above, the contribution which is relevant to heat up the neutral IGM comes from the soft X-ray band (0.5-2 keV). In this case we applied a correction factor $K_{\rm X}$ from \citet{Shen2020} (which has been found also by \citealt{Graziani2018}):   
\begin{equation}
    K_{\rm X, 0.5-2keV} = c_1\bigg(\frac{L_{\rm bol}}{10^{10}{\rm L_\odot}}\bigg)^{k_1} + c_2\bigg(\frac{L_{\rm bol}}{10^{10}{\rm L_\odot}}\bigg)^{k_2}, 
\end{equation}
with $c_1 = 5.712$ , $k_1 = -0.026$ , $c_2 = 17.67$ and $k_2 = 0.278$.
As we have already done for UV photons, we must account for the fact that some X-ray photons (probably a smaller fraction than the UV) will not escape from the galaxy where they are produced. Following \citet{Trinca2022} our reference model for X-ray obscuration is the one proposed by \citet{Ueda2014}. This correction is expressed in terms of the $\psi(L_{\rm X},z)$ parameter which represents the fraction of absorbed AGNs. It is expressed as a linear function of $\log L_{\rm X}$ within a range $\psi_{\rm min}=0.2$ and $\psi_{\rm max}=0.84$:
\begin{equation}
    \psi(L_{\rm X},z)=\min[\psi_{\rm {max}},\max[\psi_{43.75}(z)-\beta(\log L_{\rm X}-43.75), \psi_{\rm {min}}]]
    \label{eq:psi}
\end{equation}
where $\beta=0.24$ and $\psi_{43.75} (z)$ represents the absorption fraction of AGNs with $\log L_{\rm X} = 43.75$ located at $z$. This redshift dependence disappears for $z\geq2$ (which is our case) and $\psi_{43.75} \simeq 0.73$. 
The X-ray luminosity from AGNs is then $L_{\rm X,unobs} = L_{\rm X} \, [1-\psi(L_{\rm X})]$. It is worth mentioning here that, all the fitting equations for the AGNs spectra adopted in this section, have been calibrated to observations at low redshifts. Extrapolating those up to the cosmic dawn it may not be correct and lead to an overestimation of the contribution of AGNs to the radiative background. This will be better discussed in section \ref{sec:XAGNs}. 

Fig. \ref{fig:XheatRun5} shows the total soft X-ray emissivity (the one contributing to IGM heating), disentangling between the various components: Pop II/I stars (solid thin line), Pop III stars (red dashed line) and unobscured AGNs (light blue thin line). The left panel shows the results of the entire sample of \textsc{cat} galaxies and the right panel illustrates how the results are affected by removing the contribution of mini-halos.
Differently from the ionizing background, the soft X-ray emissivity from AGNs is dominant already at very high redshift ($z + 1 \sim 34$) in both panels. Indeed, a significant contribution of the bolometric X-ray emission from HMXBs falls in the hard regime (E $\geq 2$keV), while AGNs have softer X-ray spectra. Removing the contribution of mini-halos mildly affects the global evolution of the soft X-ray emission. As for the ionizing emission, the two panels in Fig. (\ref{fig:XheatRun5}) at $z + 1 > 15$ differ mostly for the red dashed curve (Pop III contribution), which is largely reduced when we are not accounting for sources hosted in mini-halos. Another interesting feature of the Pop III HMXBs emission is its "burstiness". Since we compute the X-ray luminosity to be proportional to the SFR, the scattered evolution of the dashed line in Fig. (\ref{fig:XheatRun5}) reflects the "burstiness" of Pop III star formation \citep{Furlanetto2021}. The properties of Pop III star formation predicted by \textsc{cat} will be discussed in a future work (Trinca et al. in prep.). Here we just highlight that the latest episodes of Pop III star formation occur at $z + 1 \sim 10$. The Pop II contribution (solid thin line) instead shows a smoother evolution and it becomes dominant over the AGN's contribution at $z + 1 < 10$.

\subsection{Radio background} 
\label{sec:RadioBG}

In Eq. (\ref{eq:DBT}) we have implicitly assumed that the only 21cm background against which we expect to observe the signal is the CMB. If a strong radio background with a brightness temperature $T_{\rm rad} > T_\gamma$ is already present at early times, the background brightness temperature would increase by a multiplicative factor of $1+T_{\rm rad}/T_\gamma$ \citep{Feng2018}. This means that in Eq. (\ref{eq:DBT}) we would need to substitute $T_{\rm rad}+T_\gamma$ in place of $T_\gamma$, which leads to a decrease in the differential brightness temperature (DBT). The background radio temperature determines the 21cm global signal also through Equations \ref{eq:TS} and \ref{eq:xalpha} changing the spin temperature evolution and the Lyman-$\alpha$ coupling coefficient.

As pointed out by \citet{Mittal2022}, it is extremely challenging to explain the peculiar shape of the EDGES signal with just Lyman-$\alpha$ and X-ray emission from stars. The presence of a radio background might be considered in order to explain the large amplitude of the absorption signal detected by EDGES (\citealt{Bowman2018a}). Having a deep absorption signal means that $T_{\rm s}$ is much lower than $T_{\rm rad}$, therefore, an extra radio background would make the HI spin temperature much lower than $T_{\rm rad}$ before the X-ray heating starts to be effective.

The EDGES signal could also be explained if there is a (weak and non-gravitational) interaction between baryons and dark matter (\citealt{Barkana2018}). In virtue of their earlier decoupling, dark matter particles are colder than the early cosmic gas and, if such an interaction between baryons and dark matter exists, it would provide a cooling mechanism decreasing $T_{\rm K}$ (and $T_{\rm S}$) and therefore increasing the difference between $T_{\rm S}$ and $T_{\rm rad}$. This second possibility will not be considered in this work, and in order to reproduce the EDGES signal we will estimate the radio background produced by the population of sources predicted by \textsc{cat}.

An early contribution to the radio background may come from accreting BH seeds \citep[e.g.][]{Feng2018,EwallWice2018,EwallWice2019,Mebane2020}. Following \citet{EwallWice2018}, we estimate their radio luminosity using the Eddington luminosity-scaled relation between the radio and soft X-ray (0.1 - 2.4 \, keV) luminosities for radio-quiet AGNs found by \citet{Wang2006}:
\begin{equation}\begin{split}
\log L_{\rm r}/L_{\rm Edd} &= (0.86 \pm 0.10) \, \log L_{\rm X, 0.1 - 2.4 {\rm keV}}/L_{\rm Edd} \\
& + (-5.08 \pm 0.19),
\end{split}\end{equation}
\noindent
and we boost the luminosity of radio-loud quasars (whose fraction is assumed to be $f_{\rm L} = 0.1$) by a factor $10^R=10^3$ to match the typical radio loudness found in SDSS/FIRST AGNs by \citet{Ivezic2002}.
With this approximation, the radio emissivity can be computed as \citep{EwallWice2018}\footnote{We have dropped the dependence on the duty cycle from the original expression of \citep{EwallWice2018} as the fraction of active BHs is computed self-consistently in \textsc{cat}.}:
\begin{equation}\begin{split}
    \epsilon(\nu,z)& = 1.2 \times 10^{22} \, {\rm W \, Hz}^{-1} ({\rm Mpc}/h)^{-3} \, \bigg(\frac{f_{\rm L}}{0.1}\bigg)  \, \bigg(\frac{10^{R}}{10^3}\bigg) \\
    & \times  \bigg(\frac{L_{\rm X, 0.1 - 2.4 {\rm keV}}}{0.1 \, L_{\rm Edd}}\bigg)^{0.86}
     \bigg(\frac{\rho_{\rm{BH}}(z)}{10^4 \, h^2 \, M_\odot {\rm Mpc}^{-3}}\bigg)\bigg(\frac{\nu}{1.4 \, {\rm GHz}}\bigg)^{-0.6},
    \label{eq:epsnu2}
\end{split}\end{equation}
\noindent
where $\rho_{\rm{BH}}(z)$ is the redshift dependent mass density of accreting BHs predicted by \textsc{cat} and L$_{\rm X, 0.1 - 2.4 {\rm keV}}$ is the soft X-ray luminosity computed in the previous section.
Since all the other parameters entering in Eq. (\ref{eq:epsnu2}) are not directly computed from our model, we follow \citet{Mebane2020} and we incorporate them in a single parameter $f_{\rm R}$:
\begin{equation}\begin{split}
    \epsilon(\nu,z,f_{\rm R}) & = 1.2 \times 10^{22} \, {\rm W \, Hz}^{-1} ({\rm Mpc}/h)^{-3} \, f_{\rm R} \\
     & \times \bigg(\frac{L_{\rm X, 0.1 - 2.4 {\rm keV}}}{0.1 \, L_{\rm Edd}}\bigg)^{0.86}\bigg(\frac{\rho_{\rm{BH}}(z)}{10^4 \, h^2 \, M_\odot {\rm Mpc}^{-3}}\bigg)\bigg(\frac{\nu}{1.4 \, {\rm GHz}}\bigg)^{-0.6}.
    \label{eq:epsnu3}
\end{split}\end{equation}
From the radio emissivity we can compute the specific intensity of the radio background at redshift $z$, as:
\begin{equation}
    J_\nu(z,f_{\rm R})=\frac{{\rm c}}{4\pi}(1+z)^3\int_z^\infty\frac{dz'}{(1+z'){\rm H}(z')}\epsilon\bigg(\nu\frac{1+z'}{1+z},z',f_{\rm R}\bigg),
    \label{eq:Jnurad}
\end{equation}
where for our purposes, $\nu$ will always be the rest-frequency of the 21cm line (1420.41MHz). 
The final quantity we are interested in is the brightness temperature of the radio background at $\nu= 1420.41$ MHz:
\begin{equation}
    T_{\rm{rad}}(z)=\frac{c^2 \, {\rm J}_\nu(z,f_{\rm R})}{2 \, \nu^2 \, {\rm k_B}},
    \label{eq:Trad}
\end{equation}
\noindent
that will enter in Eq. (\ref{eq:DBT}) to compute the 21cm DBT ($T_\gamma \rightarrow  T_\gamma + T_{\rm rad}$).


\begin{figure}
    \includegraphics[width=\columnwidth]{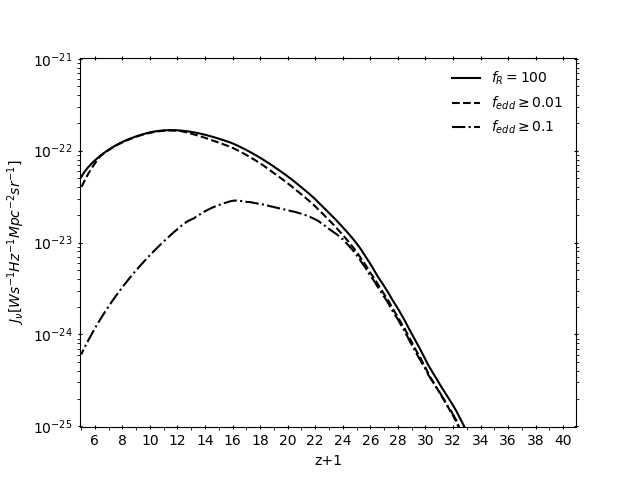}
    \caption{Redshift evolution of the specific intensity of the radio background, $J_\nu$ [W\, Hz$^{-1}$\, s$^{-1}$\, Mpc$^{-2}$\,sr$^{-1}$], from BHs, assuming $f_{\rm R} = 100$ (see text). 
    The solid, dashed and dot dashed lines show the results when all BHs and BHs accreting with an Eddington ratio of $f_{\rm edd}\geq 0.01$ and $f_{\rm edd}\geq 0.1$ are considered, as predicted by \textsc{cat}}
    \label{fig:JnuRun5_100}
\end{figure}
\begin{figure*}
    \includegraphics[width=\columnwidth]{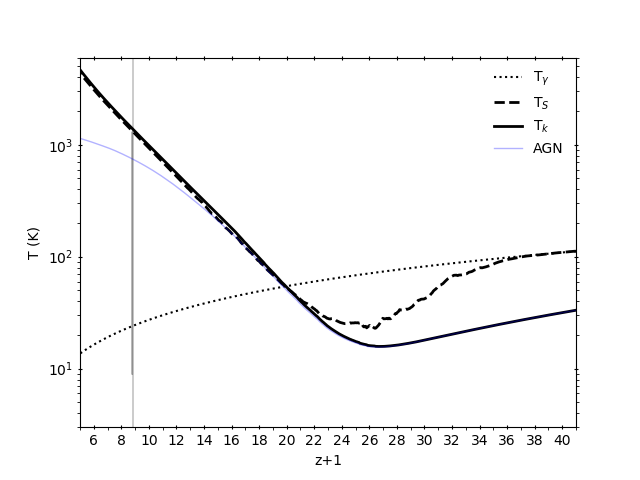}
    \includegraphics[width=\columnwidth]{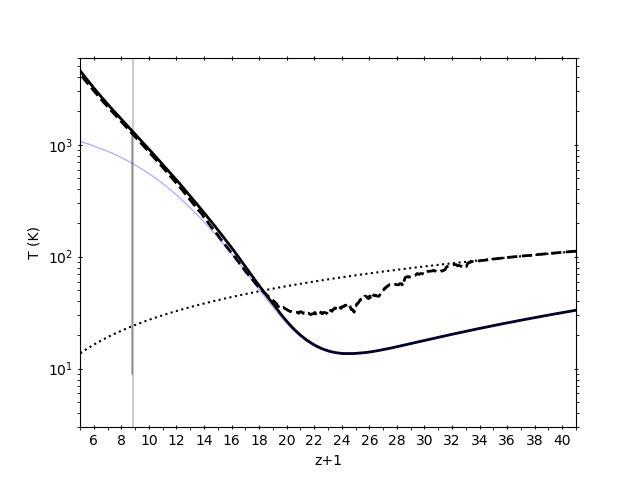}
    \caption{Redshift evolution of the spin temperature (thick dashed line) and of the kinetic temperature (thick solid line) computed starting from \textsc{cat} outputs. Dotted line shows the evolution of the CMB temperature ($\propto(1+z)$). The contribution of AGNs to heating the kinetic temperature is highlighted through the thin blue solid line. The light (dark) grey vertical bar shows the constraint on $T_{\rm K}$ at $z + 1 = 8.9$ within 68\% (95\%) confidence interval provided by 21cm power spectra measurements from HERA \citet{theheracollaboration2021hera}. As in Fig. \ref{fig:UvIonRun5} and \ref{fig:XheatRun5}, left panel considers all the galaxies predicted by \textsc{cat} while the right panel does not account for the contribution of systems hosted in molecular cooling halos.}
    \label{fig:ThermalRun5}
\end{figure*}

Fig. \ref{fig:JnuRun5_100} shows the specific intensity of the radio background as a function of redshift assuming $f_{\rm R} = 100$ (the motivation for this choice will be discussed in Section \ref{sec:EDGES}). The solid line illustrates the result when all the BHs predicted by \textsc{cat} are assumed to contribute, while the dashed and the dash-dotted lines show the radio background intensity considering only BHs that are accreting with an Eddington ratio $f_{\rm edd} \geq 0.01$ and 0.1, respectively. The motivation to disentangle between BHs accreting at different paces (consistently with \citealt{EwallWice2019}) is that the radio emissivity is expected to depend strongly on the BH accretion rate. 

The radio background predicted by our model starts to be built up as soon as the first BH seeds are formed, and rapidly increases up to $z + 1 \sim 16$. At lower redshift the growth is smoother. There is a very little difference between the solid and the dashed curve meaning that, in our model, a large fraction of BHs accretes gas with an Eddington ratio larger than 0.01. If we assume a more extreme cut at $f_{\rm edd} \geq 0.1$, the radio background drops by almost one order of magnitude at $z + 1 \leq 16$.

\section{Results} 
\label{sec:results}

Here we present the results of the 21cm global signal computed using the source properties predicted by \textsc{cat} following the procedure described in the previous section. We start with the thermal history of the IGM and the corresponding 21cm signal in Section \ref{sec:allresult} and \ref{sec:21cmAll}. In order to better disentangle the contribution of mini-halos where the first stars form, in Section \ref{sec:nominiresult} we recompute the 21cm global signal by removing the contribution of mini-halos.

\subsection{Thermal history of the IGM} 
\label{sec:allresult}

The thermal history predicted by the model when both stars and accreting BHs are considered is shown in Fig. \ref{fig:ThermalRun5}. 
Here we show the redshift evolution of the spin temperature (thick dashed line), of the CMB temperature (dotted line) and of the gas kinetic temperature (thick solid line). We also show the contribution of unobscured AGN to the kinetic temperature of the gas (blue solid line) as it is the dominant one. The two panels of Fig. \ref{fig:ThermalRun5} show the same quantities computed considering all the galaxies predicted by \textsc{cat} (left panel) and without the sources formed in mini-halos (right panel). 

At the beginning of the evolution, the IGM cools adiabatically. However, after the first sources start to form, the kinetic temperature starts to increase ($z + 1 \sim 27$, left panel) due to X-ray heating until it becomes hotter than the CMB temperature ($z + 1 \sim 20$, left panel). Thereafter $T_{\rm K}$ continues to increase reaching the maximum allowed value of $10^4$ K when hydrogen is fully ionized. We highlight that the evolution of $T_{\rm K}$ is consistent with the constraint obtained by \citet{theheracollaboration2021hera} at $z + 1 = 8.9$ (95\% confidence interval) represented by the vertical line. If we neglect the contribution of mini-halos, the evolution of $T_{\rm K}$ at low redshift is almost unaffected, while some differences can be seen at $z + 1 \geq 15$. The heating starts to be effective at $z + 1 \sim 25$ and $T_{\rm K} > T_{\gamma}$ at $z + 1 \sim 18$. The thin blue line in both panels is close to the thick black one (at $z + 1 \geq 14$ the two lines overlap), showing that the heating is largely dominated by accreting BHs at all the redshifts with a minor contribution from stars.

The spin temperature is initially coupled to the CMB temperature. Once the first structures are formed, $T_{\rm S}$ is driven toward the kinetic temperature by the Lyman-$\alpha$ photons until, at $z + 1 \sim 22$ (left panel), a tight coupling between $T_{\rm K}$ and $T_{\rm S}$ is reached and $T_{\rm S} \simeq T_{\rm K}$ thereafter. Again, the effect of not considering sources formed in molecular cooling halos is to delay the thermal evolution of the IGM; the tight coupling between $T_{\rm S}$ and $T_{\rm K}$ is reached at $z + 1 \sim 19$ when the IGM is already hotter than the CMB. While the contribution of unobscured AGN was dominant for the evolution of $T_{\rm K}$, it is negligible for the spin temperature. This is a result of the fact that AGN are highly obscured in the Lyman-$\alpha$ band. Thus they are not able to provide Lyman-$\alpha$ photons in order to couple $T_{\rm S}$ to $T_{\rm K}$. Ultimately, this tells us that this coupling must be provided by stars. 

\begin{figure*}
    \includegraphics[width=\columnwidth]{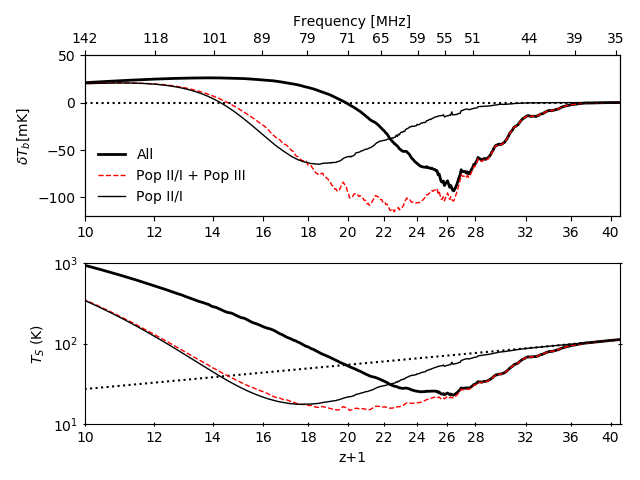}
    \includegraphics[width=\columnwidth]{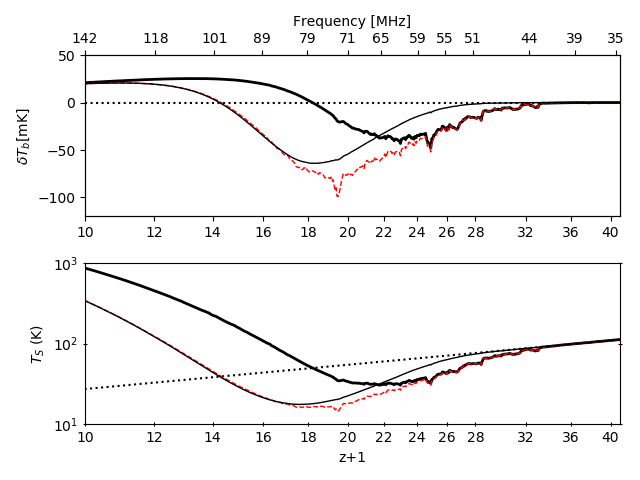}
    \caption{Upper panels: Redshift evolution of the 21cm global signal expressed as the differential brightness temperature, $\delta T_{\rm b}$ (in units of mK). This is computed considering only the contribution of Pop II stars (solid thin line), Pop II + Pop III stars (dashed red line), and all sources (solid thick line). The top x-axis shows the observed frequency of the signal. Lower panels: Redshift evolution of the spin temperature of the neutral gas computed considering only the contribution of Pop II stars, Pop II + Pop III stars and all sources using the same line coding above. These results are obtained with (left panel) and without (right panel) the contribution from the sources in mini-halos as in the previous plots.}
    \label{fig:21cmRun5}
\end{figure*}

\subsection{The 21cm global signal}
\label{sec:21cmAll}

The corresponding 21cm global signal is shown by the thick solid line in the upper left panel of Fig. \ref{fig:21cmRun5}. We find that between $z + 1 \sim 20.5$ and $z + 1 \sim 32$ (44 $\leq \nu \leq$ 69 MHz) the signal is observed in absorption, with a maximum depth of $\delta T_{\rm b} \simeq - 95$ mK at $z + 1 \sim 26.5$, which corresponds to an observed frequency of $\nu \sim 54$ MHz. This feature reflects the early coupling between the kinetic and spin temperature (see lower left panel). The early heating of the IGM due to accreting BHs (as shown in the previous section) plays an important role in determining the shape and the depth of the absorption feature. Indeed, if we neglect the contribution of AGN and leave only the stellar one (red dashed line) we predict that \textit{(i)} the absorption feature becomes wider ($\delta T_{\rm b} \leq -50$ mK until $z + 1 \sim 17$), \textit{(ii)} deeper ($\delta T_{\rm b} \sim -110$ mK at $20 \leq z + 1 \leq 26$) and \textit{(iii)} the transition from an absorption to an emission signal (which corresponds to the spin temperature to become larger than the CMB one) occurs later ($z + 1 \sim 14.5$, 97 MHz). If we subtract also Pop III stars and we let only Pop II stars contribute (solid thin line) we obtain a 21cm global signal history consistent with the "classical" evolution found by many reference studies in the literature \citep[e.g.][]{Furlanetto2006, Visbal2012, Fialkov2013a, Cohen2017, munoz2021}. In our computation, the 21cm global signal driven by only Pop II stars reaches a maximum amplitude in absorption of $\delta  T_{\rm b} \sim -65$ mK at $z + 1 \sim 18$ (78 MHz). The difference between the red dashed line and the solid thin one highlights the importance of the contribution of Pop III stars to the absorption feature as it introduces an earlier coupling between $T_{\rm S}$ and $T_{\rm K}$ (see lower left panel) which anticipates the timing and increase in the depth of the absorption feature compared to the case where only Pop II stars are present. Another effect of Pop III stars is to make the signal more "noisy". If only Pop II stars were present, the evolution of the signal would have been smoother, reflecting the "burstiness" of Pop III star formation.

Thus, as shown by the left panels of Fig. \ref{fig:21cmRun5}, the 21cm global signal encodes the properties of the sources that are present in the Universe at high redshift and their evolution from cosmic dawn to the epoch of reionization, as the various classes of sources contribute in different ways to the kinetic and spin temperature evolution of the IGM. Pop III stars are very effective at driving $T_{\rm S}$ toward $T_{\rm K}$ at high redshift, but they provide a negligible contribution to gas heating. Pop II stars are very important for the coupling and have a small impact on the heating. Finally, accreting BHs are crucial to heat the neutral hydrogen erasing most of the contribution to the signal coming from Pop II stars and leaving only the strong absorption feature caused mainly by Pop III stars. 

\begin{figure*}
    \includegraphics[width=\columnwidth]{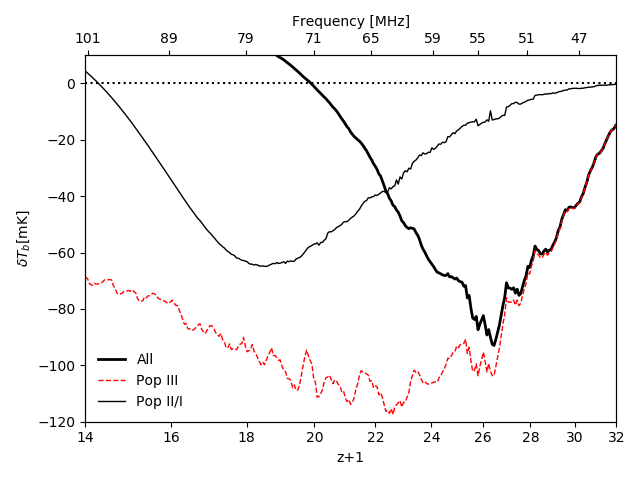}
    \includegraphics[width=\columnwidth]{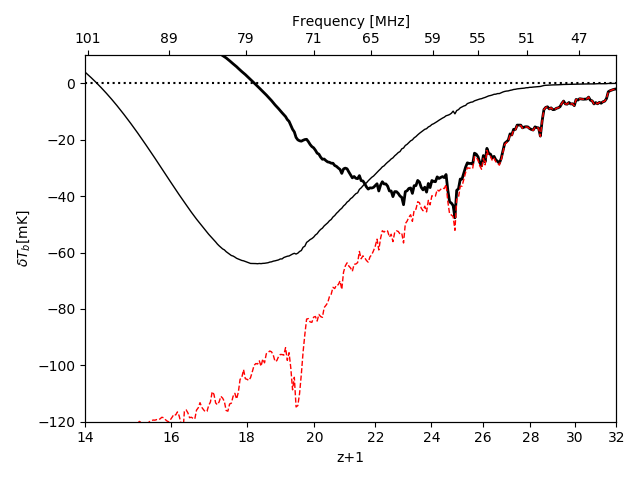}
    \caption{A zoom-in of the 21cm global signal seen in absorption at cosmic dawn when all the sources are considered (left panel) and without the contribution of sources forming in mini-halos (right panel). Here we highlight the contribution to the signal if only Pop III (Pop II/I) are present with the red dashed (black thin) line. Note that, differently from Fig. \ref{fig:21cmRun5}, the red dashed line here accounts for the Pop III contribution alone.}
    \label{fig:21cmZoomRun5}
\end{figure*}

In this work we are mostly interested in the 21cm absorption feature originating at cosmic dawn, since this provides direct evidence for the onset of Pop III star formation and may encode important information on the interplay between physical processes that regulate early star formation at $z > 17$. 
In Fig. \ref{fig:21cmZoomRun5} we show a zoom-in of the 21cm global signal between $z + 1 = 32$ and $z + 1 = 14$, highlighting separately the contributions of Pop III (red dashed line) and Pop II (thin black line) stars (we note that, differently from Fig. \ref{fig:21cmRun5} where the red dashed line accounts for the entire stellar contribution Pop III + Pop II, here the red dashed line refers to the Pop III contribution alone).
When all sources are considered (left panel), the absorption feature reaches a maximum depth $\delta T_{\rm b} \sim -95$ mK at $z + 1 \sim 26.5$ (54 MHz), with a smooth decrease starting at ($z + 1 = 34$). As we can see from the thin black and red dashed lines, the absorption signal at $z + 1 \geq 26$ is entirely determined by the early coupling caused by Pop III stars. At such high-$z$ the heating is still not effective and the IGM is not chemical enriched yet so that Pop II star formation episodes are rarer than Pop III ones. As a consequence, the coupling due to Lyman-$\alpha$ photon emission from Pop III stars is predicted to be the main contribution to the shape of the 21cm global signal at these redshifts. At lower redshifts, even if we do not consider the heating from unobscured AGNs, the stellar contribution to the heating of the IGM (coming from HMXBs as described in Section \ref{sec:XBG}) is sufficient to heat up the neutral hydrogen above the CMB temperature, suppressing the 21cm global signal (see red dashed line in Fig. \ref{fig:21cmRun5}). The zoom-in of the signal (Fig. \ref{fig:21cmZoomRun5}) shows that Pop II stars are much more effective in heating up the IGM compared to Pop III stars. The red dashed line in Fig. \ref{fig:21cmZoomRun5} is always negative ($\delta T_b \leq -70$ mK) meaning that, if we consider only Pop III stars, we are able to drive the spin temperature toward the kinetic temperature, but the latter will always be much smaller than the background radiation and thus $T_{\rm S} < T_{\gamma}$ making the signal always be in absorption. This does not happen when we compute the evolution of the 21cm global signal only from Pop II stars. The absorption feature is delayed compared to the red dashed line, but it turns into an emission signal at $z + 1 \sim 14.5$ ($\nu \sim $ 98 MHz). Ultimately this implies that Pop III stars are mostly important for the early coupling between $T_{\rm S}$ and $T_{\rm K}$ but not for the heating of the IGM, while Pop II stars have an impact on both.

\subsection{Impact of the mini-halo population}
\label{sec:nominiresult}

The previous results have been obtained considering all sources that form inside both molecular (or mini-halos) and atomic cooling halos. Here we investigate the results without the contribution of 
mini-halos, in order to see how the 21cm global signal changes. In practice, we compute the 21cm global signal considering only sources hosted by dark matter halos with $T_{\rm{vir}} \geq 10^4$\, K. It is important to stress, however, that this procedure is applied by post-processing \textsc{cat} simulations output. Hence the properties of the sources hosted in atomic cooling halos are still sensitive to the physical processes occurring in the less massive mini-halos. 

The 21cm signal evolution is shown in the upper right panel of Fig. \ref{fig:21cmRun5} (thick line). The overall profile is now quite different compared to the one obtained from the entire halo population and shown in the upper left panel of the same plot. Without the contribution of mini-halos, the absorption feature is strongly suppressed, showing a much smaller amplitude compared to the previous case ($\delta T_{\rm b} \simeq -40$ mK for $\nu \sim 55-63$). At higher frequencies, $\delta T_{\rm b}$ starts to increase until the signal vanishes at $z + 1 \sim 18.5$. This indicates that sources hosted in molecular cooling halos are crucial in achieving an early tight coupling between the spin and kinetic temperature. However, their contribution is soon cancelled out due to the effect of the Lyman-Werner feedback that stops the molecular cooling inside mini-halos preventing star formation. However, the Pop III star contribution, even if smaller, still dominates the shape of the absorption feature. This is visible when we compare the 21cm signal after removing AGN (red dashed line), with only Pop II stars (black thin line) and with both stars and AGN (black thick line). The red dashed line and the black thick one have a very similar evolution for $z + 1 > 25$ indicating that the role of accreting BHs is quite visible even when we do not consider sources in mini-halos. The signal computed only from Pop II stars instead is almost unaffected, reflecting the fact that Pop II stars mostly form in more massive atomic cooling halos. As in the previous section, the zoom-in of the absorption feature in the 21cm global signal obtained after removing mini-halos (Fig. \ref{fig:21cmZoomRun5}, right panel) shows that Pop III stars efficiently drive the coupling between $T_{\rm S}$ and $T_{\rm K}$ but do not heat up the IGM. Overall, neglecting mini-halos leads to important changes in the signal at high-$z$ when the absorption feature is predicted while at lower redshift ($z + 1 < 15$ in our simulation) star formation in mini-halos is suppressed by radiative feedback from the Lyman-Werner background, and no longer contributes to the signal. 



\subsection{Impact of a radio background}
\label{sec:radioresult}

The results shown so far consider the redshifted CMB photons as the only radio background against which the DBT is measured ($T_{\rm R} = T_\gamma$). In this subsection we show the impact of an additional radio background to the 21cm signal computed assuming the whole halo population as explained in Section \ref{sec:RadioBG}. We stress that, using \textsc{cat} outputs, we are able to compute this additional background consistently with the other radiation backgrounds that we have considered so far. The only free parameter that enters here is $f_{\rm R}$ and this can be tuned to match the amplitude of the EDGES claimed detection (\citealt{Bowman2018a}).

\begin{figure}
    \includegraphics[width=\columnwidth]{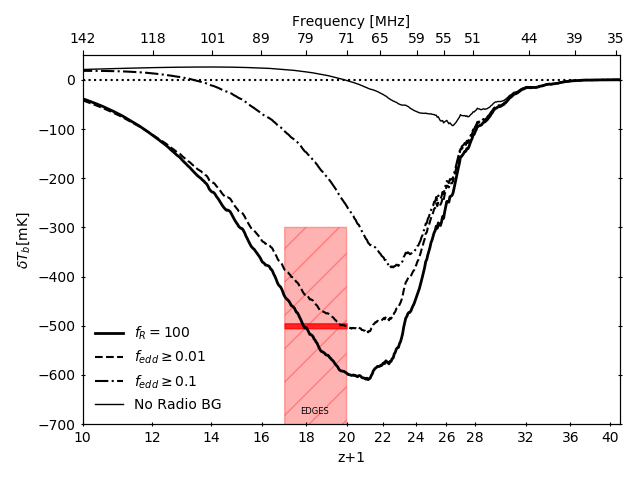}
    \caption{21cm global signal once the radio background is incorporated assuming $f_R = 100$. Solid, dashed and dot dashed lines consider a radio background computed from all BHs, BHs which are accreting with an Eddington ratio $f_{\rm edd}\geq0.01$ and $f_{\rm edd}\geq0.1$ respectively. These results are compared with the previous computation of the signal (thin line). For comparison, the EDGES claimed detection \citealt{Bowman2018a} is shown (red bar with shaded region for the error).}
    \label{fig:SignalRun5_radio100}
\end{figure}


The resulting 21cm signal is shown in Fig. \ref{fig:SignalRun5_radio100} adopting a value of $f_{\rm R} = 100$ and considering BHs that are accreting at different rates, as discussed in Section \ref{sec:RadioBG} (solid, dashed and dash-dotted lines refer to all BHs, BHs with $f_{\rm edd} \geq 0.01$ and 0.1, respectively). 
The solid thick line shows that the signal is strongly altered by the presence of a radio background. The absorption feature between $20 \leq z + 1 \leq 24$ has now an increased amplitude ($\delta T_{\rm b} \sim -600$ mK at $\nu \sim 59 - 71$ MHz). In addition, the 21cm signal is in absorption at all frequencies, with an amplitude $\delta T_{\rm b} \leq -50$ mK at $z + 1 \geq 11$. The evolution of the signal is now smoother across all the redshift steps, since the leading contribution is now the radio background from accreting BHs shown in Fig. \ref{fig:JnuRun5_100}. 

In order to match the amplitude of EDGES detection \citet{Bowman2018a}, a large value of $f_{\rm R} = 100$ is required. If we consider only BHs that are accreting with $f_{\rm edd} \geq 0.01$ (dashed line), the absorption feature at $\nu \sim 63 - 74$ MHz is mildly shallower ($\delta T_{\rm b} \sim -500 $ mK) while the remaining evolution is almost identical. Restricting the contribution to systems with $f_{\rm edd} \geq 0.1$ (dash-dotted line) leads to an absorption signal which is much weaker and potentially turning into an emission signal at $z + 1 \sim 14$ ($\nu \sim 101$).

From now on, we will take the dashed curve as the reference 21cm signal evolution when an additional radio background is considered, as this appears to be the model that best fits the amplitude (although not the timing) of the EDGES detection (see the red bar in Fig. \ref{fig:SignalRun5_radio100}). It is evident that, even considering this additional radio background we are not able to reproduce the EDGES results with our reference model. This point will be further discussed in Section \ref{sec:XAGNs} and \ref{sec:EDGES}. This choice of $f_{\rm R}$ leads to a radio background temperature consistent with the current observational constraint from \citet{theheracollaboration2021hera}: $\log_{10}(T_{\rm {rad}}/T_{\rm K}) < 1.7$ at $z + 1 = 8.9$ (95\% confidence).
\section{Discussion}
\label{sec:discussion}

The 21cm global signal that we show in the top-left panel of Fig. \ref{fig:21cmRun5} has an absorption feature peaked at $z + 1 \sim 26.5$, (54 MHz) with a depth $\delta T_{\rm b} \simeq -95$ mK followed by a sharp increase of the DBT. After removing the contribution of mini-halos (Fig. \ref{fig:21cmRun5}, top-right panel) the absorption feature in the 21cm signal evolution is strongly suppressed ($\delta T_{\rm b}$ always $ \geq -45$ mK). From $z + 1 \leq 15$ the two signals share the same trend. Analysing the separate contributions of Pop III stars, Pop II stars and accreting BHs we find that Pop III stars play a crucial role in the early coupling between $T_{\rm S}$ and $T_{\rm K}$ at $z \geq 24$, Pop II stars dominate the signal at lower redshift (when it is already in emission). Accreting BHs are not responsible for the coupling between the spin and the kinetic temperature, but are the major source of heating of the neutral IGM causing a rapid suppression of the absorption feature at high-z and an early transition from an absorption to an emission signal. In our simulations, the impact of unobscured AGNs completely washes out the impact of Pop II stars to the absorption feature as it makes the kinetic temperature hotter than the CMB background radiation before the emission of Lyman-$\alpha$ photons from Pop II stars becomes significant. The absorption feature we predict, together with many other previous studies \citep[e.g.][]{Furlanetto2006,Fialkov2013a,Cohen2017,munoz2021}, represents the main target of experiments like EDGES (\citealt{Bowman2013}) and REACH (\citealt{Cumner2022}). For this reason, from now on we will focus on this feature trying to understand what physical effects are responsible for the shape, timing and depth of the absorption feature that we have found.

\subsection{Pop III stars and 21cm absorption at z = 21-32}
\label{sec:absorption}

The main novelty of the 21cm signal model presented in this work is the inclusion of Pop III stars and BH evolution, starting from their seeding. The vast majority of current theoretical models for the 21cm global signal, identify Pop III stars with stars that are formed inside mini-halos \citep[e.g.][]{Qin2020, munoz2021} without allowing Pop III star formation in more massive atomic cooling halos. Among the few studies on the impact of Pop III stars to the 21cm global signal we mention \citet{magg2021} and \citet{Jones2022} (who focus respectively on the Pop III/II chemical transition and on the IMF of Pop III stars.) Other previous works considered the impact of Pop III stars on the radio and X-ray background in order to match the EDGES detection reported by \citet{Bowman2018a} \citep[e.g.][]{Schauer2019, Chatterjee2020, Mebane2020}.

Given the large uncertainties on the parameters that determine Pop III star formation, in this work we choose not to explore such a huge parameter space, but pick a single model of Pop III star formation (see Section \ref{sec:CAT} and Trinca et al. in prep.) and examine its impact on the 21cm signal. The choice of a Pop III star formation model also determines the formation and the evolution of the first BHs: Pop III remnants are one of the two channels through which we form BH seeds in \textsc{cat}, and therefore, depending on the characteristic IMF of Pop III stars we will end up with a different number of BH seeds. Thus, for the first time, we are able to provide a self-consistent way to estimate both the impact of Pop III stars and BHs to the 21cm signal. 

In Section \ref{sec:results} we already assessed the role of Pop III star formation in the early coupling between $T_{\rm S}$ and $T_{\rm K}$ and in the narrow 21cm absorption feature (solid thick line in Fig. \ref{fig:21cmRun5}). A detailed description  of Pop III star formation in \textsc{cat} will be presented in a forthcoming paper (Trinca et al. in prep). Here we discuss those factors that have a direct impact on the 21cm global signal.

Firstly, we notice that the absorption feature presented in this work differs from many other previous works. This is mostly caused by the fact that \textsc{cat} is able to resolve all the minihalos down to $M_{\rm halo} \sim 5\times10^5M_{\odot}$ and the SF episodes within those. This leads to a higher SFRD at $z > 25$ compared to other works (e.g. we obtain a Pop III SFRD of 0.5 dex higher than \citealt{Visbal2018} in the redshift interval 25 - 35). This small difference can also be attributed to the different MTs used. As described in Section \ref{sec:CAT}, our semi-analytical model uses \textsc{galform} which is based on the Extended Press-Schechter theory and it is tuned in order to obtain merger histories consistent with the N-body Millennium simulation (\citealt{Springel2005}). This choice leads to a higher halo number density at $z > 25$ compared to other N-body simulations ultimately impacting on the number density of star formation episodes and on the earlier coupling between $T_{\rm S}$ and $T_{\rm K}$. As the cosmic dawn is a poorly constrained epoch of the evolution of our Universe, we believe that our anticipated evolution of the 21cm global signal is not a major concern as the model that we present is self-consistent and well-anchored both to the observations at lower redshifts and, as shown in this section, to the results of the hydrodynamical simulations for Pop III star formation. Lastly, a comparison between different models of the global signal must account for the differences on how mini-halos and their relative contribution are modeled.

When Pop III star formation occurs in mini-halos, we enhance the star formation efficiency to $\rm \epsilon_{SF, PopIII} = 0.15$ in order to achieve a total stellar mass formed in each star formation episode of $M_{\rm Pop III} \geq M_{\rm min, Pop III} = 150 M_\odot$, in agreement with state-of-the-art hydro-dynamical simulations \citep{Chon2021}. We then randomly sample the assumed top-heavy Pop III IMF until we saturate $M_{\rm Pop III}$ (see \citealt{Valiante2016}, \citealt{deBennassuti2017} and \citealt{Trinca2022} for the random sampling in \textsc{cat}).
Hence, the value of $M_{\rm Pop III}$ (and thus the adopted Pop III star formation efficiency) has an impact on the masses of newly formed Pop III stars, with larger values of $M_{\rm Pop III}$ favouring a larger frequency of Pop III stars at the high-mass end of the distribution.

Their high masses and large number are the main reason why we obtain a very early coupling between $T_{\rm S}$ and $T_{\rm K}$ and thus the absorption feature at $z = 21 - 32$ as shown in the top-left panel of Fig. \ref{fig:21cmZoomRun5}. Moreover, in \textsc{cat} a significant fraction of the total number of Pop III stars are formed inside mini-halos (this is a realistic scenario as mini-halos are the first to virialize at high redshift and are more abundant than the more massive halos). For this reason, when we consider only the sources that form inside atomic cooling halos, the impact of Pop III stars on the 21cm signal is strongly reduced for $z + 1 > 21$. Since in our model this is the main contribution to the absorption feature at high-$z$, removing mini-halos strongly suppresses the absorption feature so that this signal would not be detectable even by the new generation of radio telescopes ($\delta T_{\rm b} > - 45$ mK over the entire evolution).

The modeling of the typical masses of Pop III stars also affects the evolution of high-$z$ BHs. Pop III stellar remnants (if the initial masses of Pop III stars are large enough) provide the light BH seed population which then grow via both gas accretion and mergers. Gas accretion onto these BHs provides an important source of heating of the Universe already at high redshift, causing a quick suppression of the absorption signal that around $z + 1 \sim 20.5$ and then an emission signal in our reference model. As already stressed, in our model it is the large number, rather than the masses, of BH seeds that makes the accreting BHs the main contribution to the total X-ray background. 

In conclusion, a reliable model of the 21cm global signal, and in particular of the high-redshift 21cm absorption feature, requires accurate modelling of Pop III star formation and their associated BH remnants. If Pop III stars are numerous, the 21cm signal will be characterised by a narrower and earlier absorption feature; the first effect is due to the increased X-ray emissivity of unobscured AGNs while the second is determined by the larger Lyman-$\alpha$ emissivity of Pop III stars.

\subsection{X-ray heating from accreting BHs}
\label{sec:XAGNs}

One surprising result discussed in this paper, is the role of accreting BHs in heating the IGM already at $z + 1 \sim 27$, leading to a quick suppression of the absorption signal. This is a consequence of \textit{(i)} the large number of light seeds formed after the death of Pop III stars and \textit{(ii)} 
the SED of accreting BHs adopted in this work (see section \ref{sec:XBG}). The second point comes with several caveats. The fitting functions adopted to compute the soft X-ray emission from BHs as a function of the bolometric luminosity are calibrated to observations at much lower redshift. In addition, the correction for the obscuration is calibrated on AGNs with much larger luminosities compared to the ones we have at high-$z$ (\citealt{Ueda2014},  \citealt{Shen2020}). Finally, we are implicitly assuming that all the light seeds that we are forming in \textsc{cat}, are located at the center of their hosting halo during their entire life, and so will be able to accrete the surrounding gas very efficiently. These three approximations likely lead to an overestimation of the AGN contribution to the total X-ray background. For this reason, we also compute the 21cm signal reducing the X-ray luminosity of AGNs to 10\% of the value adopted by our reference model discussed above.

\begin{figure}
    \includegraphics[width=\columnwidth]{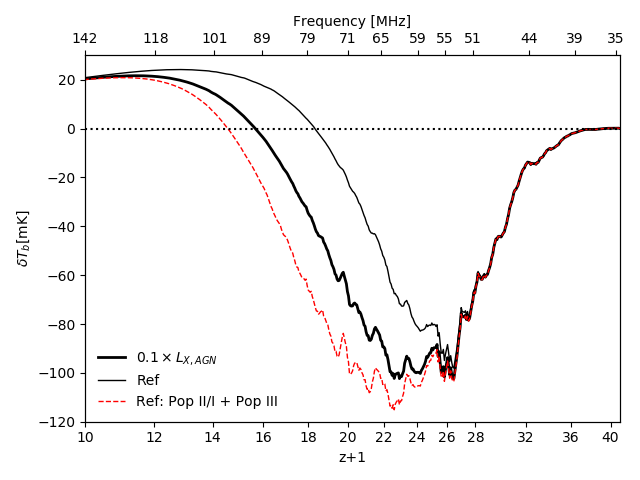}
    \caption{21cm global signal assuming that the X-ray heating coming from accreting BHs is only 10\% of the total computed in \textsc{cat} (solid thick line, see text for additional explanations). This result is compared with the previous computation of the signal accounting all sources (thin line) and only stellar sources (red dashed line).}
   \label{fig:SignalRun6_AGN}
\end{figure}


Fig. \ref{fig:SignalRun6_AGN} shows the 21cm signal computed under the approximation described above compared with our reference signal when all sources (black thin line) and only stellar sources (red dashed line) are considered. Reducing the amount of X-rays produced by unobscured AGN, leaves an imprint on the signal of a wider and slightly deeper absorption feature ($\delta T_{\rm b} \sim -100$ mK at $z + 1 \sim 22 - 26$ ($\nu \sim 54 - 64$ MHz) and delays the transition to an emission signal to $z + 1 \sim 15.5$ ($\nu \sim 91$ MHz). Nevertheless, accreting BHs would still have an impact on the signal as the black thick line still shows a faster evolution than the red dashed line. This result may suggest that the approximations done within \textsc{cat} for the estimation of the X-ray background produced by the accreting BH seeds lead to an overestimation of the total heating of the neutral hydrogen when predicting the timing of the absorption feature. Finally, we notice that the model with the reduced X-rays from accreting BHs, is consistent also with the updated constraints recently provided by HERA (\citealt{Hera2022b}).

\subsection{EDGES signal}
\label{sec:EDGES}

In the previous section we did not consider the presence of a radio background (Fig. \ref{fig:SignalRun5_radio100}); here we will discuss the consistency between our signal and the claimed detection by the EDGES collaboration \citet{Bowman2018a}.

The three key features of the absorption profile detected by EDGES are: \textit{(i)} timing: $17 \leq z + 1 \leq 20$ centered at $z + 1 \sim 18$, \textit{(ii)} depth: $\delta T_{\rm b} = -500^{+200}_{-500}$ mK and \textit{(iii)} shape: flat profile (all these features are reported in Fig. \ref{fig:SignalRun5_radio100} with a red bar and a red shaded region representing the uncertainty). Our prediction shown in Fig. \ref{fig:SignalRun5_radio100} only match the depth of the absorption profile with a maximum of $-500$ mK at $\nu \sim 63-74$ MHz. 
The most difficult of the EDGES constraints to match is the flat shape of the profile. The dashed line in Fig. \ref{fig:SignalRun5_radio100} drops quite fast at high-$z$ (but not as fast as EDGES detection), it has a broad and almost constant peak between $19 \leq z + 1 \leq 22.5$ and then the differential brightness temperature increases gently. This slow increase in the DBT is not consistent with EDGES claimed detection which instead shows a sharp increase. Finally, we note that, the effect of a radio background is to "wash out" all the "bursty" evolution originated from Pop III stars. Nevertheless, Pop III stars have a key role also in the computation of the radio background as they provide the BH seeds that, as soon as they start accreting gas, emit radiation in the radio band. Finally, we noticed that the timing of our absorption feature is predicted to be $\Delta z \sim 2$ earlier compared to EDGES. In the previous section we assessed that this is probably due to an overestimation of the X-ray background that erases the absorption feature faster. 

\begin{figure}
    \includegraphics[width=\columnwidth]{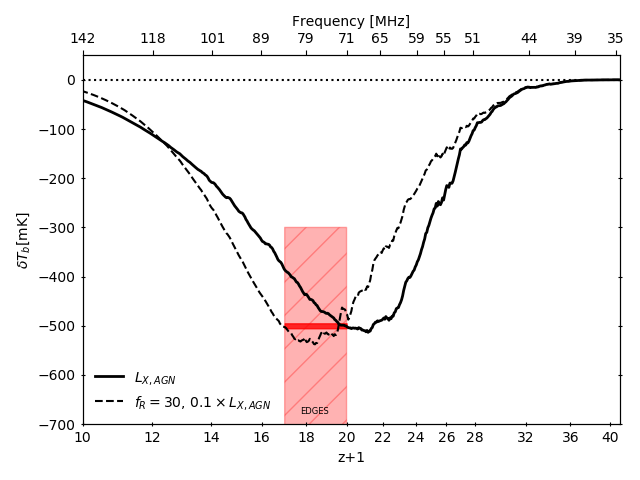}
    \caption{21cm global signal once the radio background is incorporated assuming that the X-ray heating from accreting BHs is only 10\% of the total computed in \textsc{cat} (see Section \ref{sec:XAGNs}). In order to match the EDGES prediction in this case we took $f_{\rm R} = 30$. This result (dashed line) is compared with the previous computation of the signal (solid line) in presence of a radio background. For comparison, the EDGES claimed detection \citealt{Bowman2018a} is shown (red bar with shaded region for the error).}
    \label{fig:SignalRun6_best}
\end{figure}


In Fig. \ref{fig:SignalRun6_best} we recomputed the 21cm signal in presence of a radio background with the reduced X-ray background from accreting BHs (dashed line). In this case, in order to achieve the depth of -500 mK we required $f_{\rm R} = 30$. Once we account for a smaller X-ray background from unobscured AGNs, the resulting signal is more consistent with the timing of EDGES. However, the flat profile is still not fully achieved.

In conclusion, our model is not able to correctly reproduce all features of the EDGES signal. The main issue is with regards to the peculiar flat profile of EDGES that suggests a strong X-ray heating is responsible for a quick rise in the DBT. In our model BHs are important both for the X-ray and the radio emission. However the two contributions act in the opposite direction, so, while to match the depth of the signal, we need a strong radio background, this tends to wash out the X-ray contribution so we cannot achieve a sharp drop in the DBT. These results are also consistent with what has been found in \citet{Mebane2020}.

\section{Conclusions and future perspectives} 
\label{sec:conclusions}


In this work we used \textsc{cat} a well-constrained semi-analytical model for the evolution of the first stars and BHs during Cosmic Dawn and the Epoch of Reionization to compute the evolution in redshift of the 21cm global signal disentangling the impact of each class of sources at high-$z$: Pop III stars, Pop II stars and accreting BHs. 


Our model relies on a self-consistent description of Pop III star formation and BH seeding and growth (through gas accretion and mergers) and follows  chemical evolution tracking the Pop III/Pop II transition in the high-$z$ Universe. Moreover, the model is well-anchored to observational constraints at z < 8.  The main results that we obtained are the following:
\begin{itemize}
    \item We predict a 21cm global signal with an absorption feature between $20.5 \leq z \leq 32$ (44 MHz $\leq \nu \leq 69$ MHz) with a maximum depth $\delta T_{\rm b} \sim -95$ mK at $z \sim 26.5$ ($\nu \sim 54$ MHz). The timing of this absorption feature appears at earlier epochs with respect to what has been found in many of the previous studies. However, as shown in some works that consider a large set of parameters \citep[e.g.][]{Cohen2017, Reis2021}, depending on the values chosen for the free parameters it is possible to obtain a signal with an absorption feature at very high redshift similar to the one we obtained.
    Our early absorption feature is due to the early coupling between T$_{\rm S}$ and T$_{\rm K}$ driven by Pop III star formation in mini-halos which, within \textsc{cat}, we are able to resolve as soon as they start to virialize at $z > 30$. Accreting BHs efficiently heat the IGM already at $z \leq 27$ driving a rapid increase in the DBT and reducing the amplitude of the 21cm absorption signal. Thus, Pop III stars and accreting BHs, once taken into account, effectively modify the shape of the absorption signal which has a smoother evolution than when only Pop II stars are considered.
    
    \item The modeling of Pop III star formation has an impact on the 21cm signal. Inside \textsc{cat}, in order to match the minimum stellar mass formed in a single Pop III SF episode predicted by hydrodynamical simulations (\citealt{Chon2021}), we enhanced the Pop III star formation efficiency. This increases the depth of the absorption signal as more massive stars have stronger emission in the bands relevant for the 21cm signal. More massive Pop III stars lead also to more massive BH seeds that cause a stronger heating and a faster transition from an absorption to an emission signal. We refer to a future work (Trinca et al. in prep) for a detailed discussion of the details of Pop III star formation in \textsc{cat}. 
    
    \item The impact of X-ray heating from accreting BH seeds on the 21cm global signal is not negligible. Without this contribution the absorption feature would be wider and deeper (see red dashed line in Fig. \ref{fig:21cmRun5}). However, this contribution is very uncertain at such high-$z$ as it is not observationally constrained. Our estimation of the X-ray background from BH seed accretion is likely to be an overestimate (see Section \ref{sec:XAGNs}). For this reason, we recomputed the 21cm global signal reducing the X-ray luminosity of AGN to 10\% of the value adopted in the reference model. In this case we find that the absorption feature becomes deeper and wider ($\delta T_{\rm b} \sim -100$ mK at $z + 1 \sim 22 - 26 (\nu \sim 54 - 64$ MHz) and the transition to an emission signal is delayed at $z + 1 \sim 15.5$ ($\nu\sim 91$ MHz). Despite being reduced, the impact of accreting BHs is still not negligible. The key role of early accreting BHs in heating up the IGM is strongly supported by the updated HERA constraints (\citealt{Hera2022b}).
    
    \item The environment where the first sources form can influence the evolution of the signal. We compute the 21cm line after having removed all the sources inside mini-halos in post-processing. The overall impact on the 21cm signal is to delay the timing of the absorption feature and to reduce its depth. The resulting absorption feature is now almost completely erased having a depth $\delta T_{b} > -45$ mK (such a shallow signal would not be observable even by SKA-low). This evolution is caused by the strong decrease of the Pop III stellar contribution to the Lyman-$\alpha$ coupling (at $z \sim 25$ the depth of the signal is a factor of two lower than in the reference model). We note that, if the signal is computed only from Pop II stars hosted in atomic cooling halos, the absorption feature is more consistent with previous studies, showing that the modeling of the 21cm global signal is very sensitive to the details of both the properties and the environments of the first sources. 
    
    \item Finally, we considered the impact on the signal of an additional radio background coming from accreting BH seeds as proposed by \citet{Feng2018,EwallWice2018}. We modeled the radio emission using the free parameter f$_{\rm R}$ that boosts it and we chose f$_{\rm R} = 100$ in order to match the depth of the EDGES claimed detection \citet{Bowman2018a}. The radio background that we obtain is consistent with the observational constraints at $z \sim 8$ provided by \citet{theheracollaboration2021hera} ($\log_{10}(T_{\rm {rad}}/T_{\rm K}) < 1.7$ with 95\% confidence). Under these conditions, we find an absorption signal which is at higher redshift compared to EDGES by $\Delta z \sim 2$ and a profile that does not quite match the flat profile of EDGES. If instead we consider the X-ray background from accreting BHs reduced by one order of magnitude, we can match both the depth and the timing (but not the profile) of EDGES assuming $f_{\rm R} = 30$. 
    \end{itemize}
    
Our findings provide additional evidences of the rich amount of information encoded in the 21cm global signal from cosmic dawn, whose detection would provide fundamental constraints on the nature of the first sources and on the physical processes
that regulate early star formation.

\section*{Acknowledgements}
We thank Y. Qin and B. Greig for the insightful discussion on the X-ray background that lead to an improvement of this work. This research was supported by the Australian Research Council Centre of Excellence for All Sky Astrophysics in 3 Dimensions (ASTRO 3D, project \#CE170100013). We acknowledge support from the Amaldi Research Center funded by the MIUR program \lq \lq Dipartimento di Eccellenza \rq \rq (CUP:B81I18001170001) and from the INFN TEONGRAV specific initiative.

\section*{Data Availability}

The simulated data underlying this article will be shared on reasonable request to the corresponding author.



\bibliographystyle{21cm}
\bibliography{21cm} 







\bsp
\label{lastpage}
\end{document}